\documentstyle[11pt]{article}

\begin{document}
\large

\def\lsim{\mathrel{\rlap{\lower3pt\hbox{\hskip0pt$\sim$}}
    \raise1pt\hbox{$<$}}}         %less than or approx. symbol
\def\gsim{\mathrel{\rlap{\lower4pt\hbox{\hskip1pt$\sim$}}
    \raise1pt\hbox{$>$}}}         %greater than or approx. symbol
\def\dblint{\mathop{\rlap{\hbox{$\displaystyle\!\int\!\!\!\!\!\int$}
}
    \hbox{$\bigcirc$}}}
\def\ut#1{$\underline{\smash{\vphantom{y}\hbox{#1}}}$}

\newcommand{\beq}{\begin{equation}}
\newcommand{\eeq}{\end{equation}}
\newcommand{\dem}{\Delta M_{\mbox{B-M}}}
\newcommand{\dega}{\Delta \Gamma_{\mbox{B-M}}}

\newcommand{\ind}[1]{_{\begin{small}\mbox{#1}\end{small}}}
\newcommand{\tind}[1]{^{\begin{small}\mbox{#1}\end{small}}}

\newcommand{\WA}{{\em WA}}
\newcommand{\SM}{Standard Model }
\newcommand{\QCD}{{\em QCD }}
\newcommand{\KM}{{\em KM }}
\newcommand{\hscale}{\mu\ind{hadr}}
\newcommand{\sG}{i\sigma G}

\newcommand{\MS}{\overline{\mbox{MS}}}
\newcommand{\pole}{\mbox{pole}}
\newcommand{\aver}[1]{\langle #1\rangle}

\newcommand{\appa}{\mbox{\ae}}
\newcommand{\CP}{{\em CP } }
\newcommand{\fy}{\varphi}
\newcommand{\hi}{\chi}
\newcommand{\al}{\alpha}
\newcommand{\as}{\alpha_s}
\newcommand{\gf}{\gamma_5}
\newcommand{\gm}{\gamma_\mu}
\newcommand{\gn}{\gamma_\nu}
\newcommand{\be}{\beta}
\newcommand{\ga}{\gamma}
\newcommand{\de}{\delta}
\renewcommand{\Im}{\mbox{Im}\,}
\renewcommand{\Re}{\mbox{Re}\,}
\newcommand{\GeV}{\,\mbox{GeV}}
\newcommand{\MeV}{\,\mbox{MeV}}
\newcommand{\matel}[3]{\langle #1|#2|#3\rangle}
\newcommand{\state}[1]{|#1\rangle}
\newcommand{\ra}{\rightarrow}
\newcommand{\ve}[1]{\vec{\bf #1}}

\newcommand{\rhs}{{\em rhs}}
\newcommand{\pp}{\langle \ve{p}^2 \rangle}

\newcommand{\BR}{\,\mbox{BR}}
\newcommand{\La}{\overline{\Lambda}}
\newcommand{\Lam}{\Lambda\ind{QCD}}

\newcommand{\np}{nonperturbative}
\newcommand{\re}[1]{ref.~\cite{#1}}

\begin{flushright}
\large{
TPI-MINN-93/60-T\\
UMN-TH-1231-93\\
CERN-TH.7129/93\\
UND-HEP-93-BIG\hspace*{0.1em}08}\\
December 30, 1993\\
\end{flushright}
\vspace{.4cm}
\begin{center} \LARGE
{\bf On the Motion of Heavy Quarks inside Hadrons:
Universal Distributions
and Inclusive Decays }
\end{center}
\vspace*{.4cm}
\begin{center} \Large
I.I. Bigi\\
{\normalsize\it TH Division, CERN, CH-1211 Geneva 23,
Switzerland \footnote{During the academic year 1993/94}}\\
and\\
{\normalsize\it Dept. of Physics,
University of Notre Dame du
Lac, Notre Dame, IN 46556, U.S.A.\footnote{Permanent address}}
\\{\normalsize \it e-mail address: VXCERN::IBIGI, BIGI@UNDHEP}
\vspace{.4cm}
\\
M.A. Shifman\\
{\normalsize\it  Theoretical Physics Institute, Univ. of Minnesota,
Minneapolis, MN 55455}
\\{\normalsize \it e-mail address: SHIFMAN@VX.CIS.UMN.EDU}
\vspace{.4cm}
\\
N.G. Uraltsev\\
{\normalsize\it TH Division, CERN, CH-1211 Geneva 23,
Switzerland}\\
and\\
{\normalsize \it St. Petersburg Nuclear Physics Institute,
Gatchina, St. Petersburg 188350, Russia $^2$}
\\{\normalsize \it e-mail
address:  VXCERN::URALTSEV, URALTSEV@LNPI.SPB.SU}
\vspace{.4cm}
\\
A.I. Vainshtein\\
{\normalsize\it  Theoretical Physics Institute, Univ. of Minnesota,
Minneapolis, MN 55455}\\and\\
{\normalsize\it Budker Institute of Nuclear Physics, Novosibirsk
630090, Russia}
\\{\normalsize \it e-mail address: VAINSHTE@VX.CIS.UMN.EDU}
\end{center}
\thispagestyle{empty} \vspace{.4cm}
\newpage

\centerline{\Large\bf Abstract}
\vspace{.4cm}

In previous papers we have pointed out that there exists a
QCD analog of the phenomenological concept of the so-called  Fermi
motion
for the heavy quark inside a hadron. Here we show
in a more detailed way how this comes about and we analyze the
limitations of this concept. Non-perturbative as well as
perturbative aspects are included. We emphasize both the
similarities
and the differences to the well-known treatment of deep inelastic
lepton-nucleon scattering. We derive a model-independent {\em
lower}
bound on the kinetic energy of the heavy quark inside the hadron.

\newpage
\addtocounter{footnote}{-2}

\section{Introduction}

The theoretical and experimental study of
deep inelastic lepton-nucleon scattering
(hereafter referred to as DIS) was instrumental in
developing QCD. Inclusive weak
decays of heavy flavors -- in particular semileptonic decays --
are intimately related to DIS via channel crossing.
It is then quite surprising to note
that for a long time there were hardly any attempts
to treat charm and beauty decays in QCD proper rather than
within some phenomenological models. The first systematic
attempt was undertaken in ref. \cite{1};  some general
observations were made \cite{2} in the context of Heavy
Quark Effective Theory (HQET) \cite{3}. The real
breakthrough occurred recently \cite{4,5,6} once a previous
stumbling block had been removed \cite{BU}.
Through an expansion in inverse powers of the
heavy quark mass $m_Q$ one has obtained
a consistent treatment based on QCD.

One finds, not surprisingly, that the inclusive decays of heavy
flavor hadrons are described by the decay of free heavy flavor
quarks in the limit $m_Q\ra \infty$. (Generically we will denote
the heavy quark by $Q$ and the heavy flavor hadron by $H_Q$).
For
finite quark masses there arise pre-asymptotic non-perturbative
corrections. A very obvious example is the following: the mass
of $H_Q$ exceeds that of $Q$; therefore there is a kinematical
regime in the $H_Q$ decay spectrum that cannot be
populated in the free quark decays. At first sight there would seem
to be a clear conflict between the observation and the
$1/m_Q$ expansion. Upon further reflection one realizes that it
is the motion of the heavy quark $Q$ inside the hadron $H_Q$
that will close this gap in the theoretical spectrum. In this paper
we will give a more rigorous justification of this intuitive
picture, both in its perturbative as well as non-perturbative
aspects.
Our treatment
of the momentum distribution of the heavy quark
is based directly on QCD without recourse to phenomenological
assumptions. We will introduce distribution
functions that are universal in the sense that they will
determine bound state effects in all inclusive transitions of a given
heavy hadron.  For instance, the very same function will define both
the shape of the photon line in $B \rightarrow \gamma X_s$
transition and the lepton energy spectrum near the end-point in the
$B \rightarrow l\bar\nu_l X_u$ semileptonic decays.

Historically Bjorken was the first to discuss the effects due to the
heavy quark motion inside the heavy hadrons in the problem of the
heavy quark fragmentation \cite{bjorken}. In application to the
heavy hadron decays similar ideas were laid in the basis of the
well-known AC$^2$M$^2$ model \cite{ACM} engineered as a simple
non-relativistic model of the heavy quark motion (which was
referred to as the `Fermi motion'). This model is now extensively
used in the analysis of the lepton energy spectrum in the
semi-leptonic decays. Suffice it to mention that the published  results
on the semi-leptonic branching ratios in $B$ mesons  rely on this
model to this or that extent.

 Our approach is  quantitative and systematic. It reveals,
in particular, that some aspects of the AC$^2$M$^2$ model
\cite{ACM} can not be reconciled with QCD.
The main observation -- that the QCD analog of a phenomenologically
introduced Fermi motion of the heavy quark inside the heavy
hadron exists
in QCD --
has been already stated and used to interpret the results in our
previous
paper on lepton spectra in inclusive heavy flavor decays
\cite{5}. Here we will present them in a more detailed and
complete form.
In the meantime papers \cite{randal2,neubert} have appeared
treating these
and related phenomena from close positions. In many instances,
where they overlap with our analysis,
there is a general agreement, yet we go beyond these treatments.
In presenting our approach we focus, mainly for pedagogical reasons,
on a simple example of the $b\ra s\gamma$ transition. The results
obtained, however, are of general validity. In particular, we adapt
the general theory to include the case of the differential distributions
in the semi-leptonic decays.

The paper is organized as follows: in Sect. 2 we
present a simplified analysis of  $b\ra s+\gamma$; basic
non-perturbative  elements of our approach are discussed and a
heavy quark distribution function is  introduced. In Sect. 3 we
treat radiative and semileptonic beauty decays in QCD
assuming that the produced quark is massless. Sect. 4 is devoted to
the case of the massive  (non-relativistic) final quark. In Sect.  5 we
summarize the impact of
radiative QCD corrections. In Sect. 6 we present a few comments on
the literature. In Sect. 7 the {\em lower} bound on the kinetic
energy of the heavy quark is
obtained along with some speculations regarding the upper bound.
Sect. 8 summarizes our results.

\section{A Toy Model for $b\ra s+\gamma$}

The phenomenon of the Fermi motion manifests itself in its
purest form in $b\ra s+\gamma$ transitions. For at the
level of the free quark decay the photon energy is fixed by
two-body kinematics to be
\beq
E_\gamma^{(0)}=\frac{m_b}{2}
\label{1}
\eeq
where the strange quark mass $m_s$ has been neglected. At the
same time
in the inclusive decays of the $B$ meson, $B \ra \gamma
X_s$, there is a spread in the  photon energy up to the
kinematical boundary:
\beq
E_\gamma^{max}= \frac{M_B}{2}.
\label{2}
\eeq
As mentioned in the Introduction, the motion of the $b$ quark
inside the $B$ meson will smear out the infinitely narrow
photon line in the free $b$ quark decay. To obtain a quantitative
treatment of this phenomenon we first `peel off' all
inessential complications like the spin of the quarks and that of the
photon; i.e. we consider a toy example where the heavy quark $Q$ is
a  scalar and there is a real scalar
field $\phi$ coupling the heavy quark field $Q$
to the light scalar quark field $q$:
\beq
 {\cal L}_\phi = h \bar{Q} \phi q \; + \; {\rm h.c.} \; ,
\label{3}
\eeq
where $h$ is the coupling constant, $\bar{Q}=Q^{\dagger}$ and the
masses of
the scalar field $\phi$
and the quark $q$ are  set to zero. Later on we will briefly discuss
the case $m_q\neq 0$. The field $\phi$ carries color charge zero;
the
reaction $Q\ra q+\phi$ is thus a toy model for the radiative
decays. It will allow us to focus on the non-perturbative
effects; the conceptually
quite relevant perturbative corrections will be
addressed in Sect. 5.

The total  width for the free quark decay
$Q\rightarrow q +\phi$ is given by the following expression:
\beq
\Gamma (Q\rightarrow q\phi ) =\frac{h^2}{16\pi \,m_Q}\equiv
\Gamma_0.
\label{5}
\eeq
The energy spectrum of $\phi$ in this limiting case
is an infinitely narrow line, see Fig. 1,
\beq
\frac{d\Gamma}{dE} = \Gamma_0\delta (E -\frac{m_Q}{2}
).
\label{6}
\eeq
The strong interaction will smear this line over a
finite interval of order
\beq
\La = M_{H_Q} - m_Q .
\label{7}
\eeq
and produce a picture of the type depicted in Fig. 2.
(The interval of energies between $m_Q/2$ and $M_{H_Q}/2$ will be
referred
to as {\em window} throughout this paper. The window and the
adjacent
domain below $m_Q/2$ whose size is $\sim\La$ will be
generically
called the end-point domain.)

Introducing the transition operator defined as
\beq
\hat T = i\int d^4 x \,{\rm e}^{-iqx} T\{ \bar Q(x) q(x) \, , \,\bar q (0)
Q (0)\}
\label{8}
\eeq
allows us a rigorous treatment of the spectrum since
the $\phi$ spectrum in the inclusive decay is obtained from
$\hat T$ in the following way:
\beq
\frac{d\Gamma}{dE} = \frac{h^2 E}{4\pi^2M_{H_Q}}
 \Im \langle H_Q|\hat T |H_Q \rangle .
\label{9}
\eeq
One applies the  Wilson operator product
expansion \cite{wilson} (OPE) to express the non-local
operator $\hat T$ through an infinite series
of local operators with calculable coefficients.
Underlying this ansatz are certain assumptions
concerning analyticity which establish duality between the
quark-gluon and hadronic descriptions in eq. (\ref{9}); for more
details see  \cite{2,5,6}.

To construct the OPE explicitly we consider the transition
operator $\hat T$ first in the tree approximation (Fig. 3) where it
takes the simple form \cite{6} :
\beq
\hat T = -\int d^4 x (x|\bar Q(X)\frac{1}{({\cal P}-q)^2}Q(X)|0)
\label{10}
\eeq
We have used here the Schwinger (background field) technique
\cite{schwinger} and the corresponding notations;
i.e., the coordinate and momentum operators $X$ and ${\cal P}$
are defined by
\beq
{\cal P}_\mu = p_\mu + A_\mu (X) , \,\,\, A_\mu =gT^aA_\mu^a;
\label{11}
\eeq
\beq
[p_\mu X_\nu ] =ig_{\mu\nu}, \,\,\, [p_\mu p_\nu ] = [X_\mu X_\nu
] = 0 ,
\label{12}
\eeq
while the states $|x)$ are the eigenstates of $X_\mu$, $X_\mu |x) =
x_\mu |x)$.

It is obvious that if the quark $Q$ considered is very heavy its
momentum contains a large mechanical part,
\beq
(P_0)_\mu  =m_Qv_\mu ,
\label{13}
\eeq
where $v_\mu$ is the four-velocity of the hadron $H_Q$, $v_\mu =
(p_{H_Q})_\mu /M_{H_Q}$. In other words, in the $x$ dependence of
$Q(x)$ the mechanical part can be singled out,
$$Q(x)\propto
{\rm e}^{-iP_0x}$$
and, correspondingly, one can proceed from the operator ${\cal
P}_\mu$ to $\pi_\mu$ presenting the difference between ${\cal
P}_\mu$ and the mechanical part $(P_0)_\mu$,
\beq
{\cal P}_\mu = (P_0)_\mu + \pi_\mu .
\label{14}
\eeq
Assembling all these definitions it is not difficult to rewrite the
transition operator (\ref{10}) in the form
\beq
\hat T = - \int d^4 x (x|\bar Q\frac{1}{(P_0 - q +\pi )^2}Q|0),
\label{15}
\eeq
which is
particularly suitable for constructing OPE by expanding eq. (\ref{15})
in the powers of $\pi$.
To zero-th order, when one neglects $\pi$ altogether, one gets
\beq
{\hat T}^{(0)} = - \frac{1}{k^2} \, \bar Q(0) Q(0) \; ,
\label{16}
\eeq
where
\beq
k=P_0 - q .
\label{17}
\eeq
Substituting this result in the general expression (\ref{9}) one
recovers the free quark prediction for $d\Gamma /dE $ ,
eq. (\ref{6}), for $m_Q\ra \infty$ since
$\langle H_Q|\bar Q Q|H_Q\rangle =1$
holds in that limit.

All pre-asymptotic corrections in the matrix element
$<H_Q|\bar Q Q|H_Q> $ to order $1/m_Q^2$
are reducible to the quantities $\bar \Lambda$
and $\langle H_Q| \bar Q {\vec\pi}^2Q|H_Q \rangle $
\footnote{There is no analog of the chromomagnetic
operator for scalar quarks.}. To
see that this is  indeed the case let us first consider the heavy quark
current,
$\bar Q \, i {\stackrel{\leftrightarrow}{D}}_\mu Q$, whose diagonal
matrix
element is fixed by conservation of current. In the rest frame of
$H_Q$
we have:
\beq
\frac{1}{2M_{H_Q}} \langle H_Q|\bar Q \,i
{\stackrel{\leftrightarrow}{D}}_0 Q
|H_Q\rangle= 1 \; \; .
\label{18}
\eeq
If we next use the decomposition
(\ref{14}) we arrive at the following relation
$$
1= \frac{1}{M_{H_Q}} \langle H_Q|m_Q\bar Q Q +\bar Q \pi_0 Q|H_Q
\rangle =
$$
\beq
 \frac{m_Q}{M_{H_Q}} \langle H_Q|\bar Q Q|H_Q\rangle
+\frac{1}{2M_{H_Q}m_Q}
\langle H_Q|\bar Q {\vec\pi}^2Q|H_Q \rangle,
\label{19}
\eeq
where the second line is due to the equation of motion. Eq. (\ref{19})
implies, in turn, that
\beq
 \frac{1}{2M_{H_Q}}\langle H_Q|\bar Q Q|H_Q \rangle =
\frac{1}{2m_Q} \left( 1
-\frac{1}{2m_Q M_{H_Q}} \langle H_Q|\bar Q {\vec\pi}^2Q|H_Q
\rangle \right) .
\label{20}
\eeq
The differential and inclusive widths are determined by the very
same combination, $(2M_{H_Q})^{-1}\langle H_Q|\bar Q
Q|H_Q\rangle$. Therefore, the results (\ref{5}), (\ref{6}) of the
leading approximation remain intact\footnote{The $1/m_Q$
corrections at the
intermediate stage appear due to relativistic normalization; in
principle
the calculations can be reformulated in such a way that they do not
appear at
all.}
at the level of $1/m_Q$ and get
corrected only at the level $1/m_Q^2$. The $1/m_Q^2$ correction in
$(2M_{H_Q})^{-1}\langle H_Q|\bar Q  Q| H_Q\rangle$ is not the only
source of the  $1/m_Q^2$ corrections in the inclusive widths,
however.

Let us consider now terms ${\cal O}(\pi)$ and ${\cal O}
(\pi^2)$ in the expansion of eq.
(\ref{15}). They are
\beq
\hat T^{(1)} =  \frac{2 k_{\mu}}{k^4} \bar Q \pi_{\mu} Q ,
\label{21}
\eeq
\beq
\hat T^{(2)} =  \frac {1}{k^4} \bar Q \pi^2 Q -
\frac{4k_{\mu}k_{\nu}}{k^6}
\bar Q \pi_{\mu} \pi_{\nu} Q .
\label{22}
\eeq

One should keep in mind that the operators in $\hat T^{(1)}$ and
$\hat T^{(2)}$ are to be averaged over $H_Q$, and we assume that
this hadron  is spinless or, if it has a non-vanishing spin, averaging
over the spin is implied. Moreover, we will also use the equation of
motion $\pi_0 Q = ({\vec\pi}^2/2m_Q) Q$. Then
\beq
\langle H_Q|\hat T^{(1)}+\hat T^{(2)}|H_Q\rangle =
\langle H_Q|\bar Q {\vec\pi}^2 Q|H_Q\rangle
\frac{1}{k^4} \left( \frac{k_0}{m_Q} -1-\frac{4}{3}
\frac{{\vec k}^2}{k^2}\right) .
\label{23}
\eeq
In $\hat T^{(2)}$  we neglected the term $\propto\pi_0^2$ which is
of the higher order in $1/m_Q$.

Now, what corrections to the energy spectrum of $\phi$, eq. (\ref{6}),
stem from the expansion of the transition operator we have just
derived? Using the general relation (\ref{9})  we readily obtain
\beq
\Delta\frac{d\Gamma}{dE} = \Gamma_0
\langle H_Q|\bar Q {\vec\pi}^2 Q|H_Q\rangle
\frac{E^2}{M_{H_Q}m_Q^2}
\left[ \delta ' (E-\frac{m_Q}{2}) + \frac{E}{3}\delta ''(E-\frac{m_Q}{2})
\right] .
\label{24}
\eeq
If in the leading approximation the spectrum was just a delta
function, the corrections are even more singular!  This is no surprise,
of course. As was mentioned above the width of the $\phi$ line in
the transition $H_Q\rightarrow \phi X_q$ is of order $\La$.
We expand in the powers of $\La /m_Q$; hence we must
expect the enhancement of the singularities in each successive order.
It is clear that to describe the shape of the line one needs to sum up
the infinite number of terms in this expansion. We will return to the
issue of summation later on. Here let us notice that the corrections to
integral quantities, such as the total width or the average energy, are
properly given by integrating eq. (\ref{24}).  In particular, it is not
difficult to see by explicit integration  that the correction to the total
width due to (\ref{24}) vanishes. Then the  difference between
$\Gamma$ and the free quark width $\Gamma_0$ comes from the
matrix element of $\bar Q Q$, see eq. (\ref{20}),
\beq
\Gamma = \Gamma_0 \left( 1-
\frac{\langle{\vec\pi}^2\rangle}{2m_Q^2} + ... \right).
\label{A}
\eeq
The correction in the brackets has a transparent physical meaning, it
reflects time dilation for the moving quark. In other words one can
say that the $1/m_Q$ in $\Gamma_0$ is merely substituted
by $\langle 1/E_Q\rangle $ in the formula for the decay
probability\footnote{Only the
kinetic energy enters here; if one identifies $\pi$ with usual
momenta -- what is assumed here -- the Coulomb potential energy
drops
out and $E_Q$ differs from $m_Q$ only by the kinetic energy.}.

As for the average energy, eq. (\ref{24}) implies that
\beq
\langle E- \frac{m_Q}{2}\rangle = \frac{1}{4M_{H_Q}}
\langle H_Q|\bar Q{\vec\pi}^2 Q|H_Q\rangle .
\label{25}
\eeq
Moreover, the dispersion is also calculable,
\beq
\langle (E- \frac{m_Q}{2})^2\rangle =
\frac{m_Q}{12M_{H_Q}}\langle H_Q|\bar Q{\vec\pi}^2 Q|H_Q\rangle .
\label{26}
\eeq
The last equation shows that the line width has a natural hadronic
size of order of the characteristic spatial momentum of the heavy
quark inside the hadron (i.e. of order
$\La$). At the same time, the shift of the center of the line,
eq. (\ref{25}), is small -- of order of $\langle {\vec\pi}^2\rangle
/m_Q$. To find higher moments of $E-(m_Q/2)$ one should proceed
to consideration of operators of higher dimension.

The vanishing of the $\bar Q
{\vec\pi}^2 Q$ correction to the total width due to eq. (\ref{24}) can
be simply understood on general
grounds. Indeed, this operator is not Lorentz invariant and clearly
can not appear by itself in the OPE for the total width which is the
Lorentz scalar. As for eq. (\ref{20}) where the same
operator enters we used there the mere substitution of Lorentz
scalar
$\bar Q Q$ by a sum of two operators,
$\bar Q \,i {\stackrel{\leftrightarrow}{D}}_0 Q $ and $\bar Q
{\vec\pi}^2 Q$; both are not Lorentz scalars.
However, operators with non-zero Lorentz spin contribute to the
energy
distribution, their Lorentz indices are contracted with the 4-
momentum
of the $\phi$ particle.

The first Lorentz scalar operator after $\bar Q Q$ is the operator
$\bar Q  Q \bar q q$ whose dimension is four -- two units higher
than that of $\bar Q Q$. The diagram giving rise to this operator is
depicted in Fig. 4; it contains a hard gluon exchange , i.e. extra
$\alpha_s (m_Q)$ in the corresponding coefficient. The operator
$\bar Q  Q \bar q q$ produces a correction $\propto
\alpha_s (m_Q)/m_Q^2$ in the total width and in the spectrum which
will not be considered in this work.

The remark above concludes a summary of those results which have
been already discussed previously in other contexts, see \cite{4} --
\cite{6}. We proceed now to the issue of the line shape. As was
mentioned above, the question of how the delta function line of the
parton  approximation is smeared into a physical finite size line
requires summation of all orders in $\pi$ in the expansion of eq.
(\ref{15}).  The formal series stemming from (\ref{15}) is
\beq
\hat T = -\frac{1}{k^2}\sum_{n=0}^\infty
\bar Q \left( -\frac{2k\pi +\pi^2}{k^2}\right)^n Q .
\label{27}
\eeq
To evaluate the relative roles of different terms in eq. (\ref{27}) one
should keep in mind that the characteristic values of the momentum
$k=P_0 -q$ in the domain of interest are
$$
k_0\sim |\vec k|\sim m_Q/2 ,\,\,\, k^2\sim m_Q\La .
$$
If the first estimate is quite obvious, the second one, probably, calls
for a comment. Kinematically $-k^2 =2m_Q(E-(m_Q/2))$ and inside
the line $|E-(m_Q/2)|\sim \La$. In particular, on the
positive side the maximal accessible value of $E-(m_Q/2)$ is
$(M_{H_Q}-m_Q)/2 =\La /2$.

It is clear then that the $\pi^2$ terms in eq. (\ref{27}) can be
systematically neglected compared to $2k\pi$. Indeed, all terms
$(2k\pi /k^2)^n$ are of the order of unity in the domain under
consideration. At the same time, insertion of every additional
$\pi^2 /k^2$ factor
leads to a suppression of order $\La /m_Q$. These
insertions  become important only at the subleading level, a question
that will not be touched upon here.

Thus, keeping only those terms in the series that do not vanish in the
limit $m_Q\rightarrow\infty$ we get the following expansion for
$\hat T$:
\beq
\hat T =
-\frac{1}{k^2}\sum_{n=0}^\infty \left( -\frac{2}{k^2}\right)^n
k^{\mu_1}...k^{\mu_n}\left( \bar Q \pi_{\mu_1}...\pi_{\mu_n}Q
-\, \mbox {traces}\right) .
\label{28}
\eeq
This expansion is very close to the expression quite standard in the
OPE analysis of deep inelastic scattering. The terms kept are analogs
of the twist two operators appearing in the latter case; those with
$\pi^2$ which we neglected represent higher-twist effects. Only
symmetric traceless operators of the leading twist are relevant to the
approximation accepted in this work. In other words, the theory of
the line shape in the heavy quark decays developed here is similar
to the theory of the structure functions in deep inelastic scattering
with the power $1/Q$ corrections discarded.  We will be returning to
this parallel with deep inelastic scattering more than once.

After the transition operator $\hat T$ is built the next step is the
averaging of eq. (\ref{28}) over the hadronic state $H_Q$. Using only
the general arguments of Lorentz covariance one can write
\beq
\langle H_Q|\bar Q \pi_{\mu_1}...\pi_{\mu_n}Q -\, \mbox {traces}
|H_Q\rangle
= a_n{\La}^n (v_{\mu_1}...v_{\mu_n} -\, \mbox {traces}).
\label{29}
\eeq
Notice that we single out the dimensionful factor ${\La}^n$,
a natural scale for $\langle\pi^n\rangle$, so that all coefficients
$a_n$
are dimensionless numbers of order one (if $n$ is not parametrically
large); in principle any other physical scale of order $\La$ can be
used
instead. With these definitions we
readily arrive at
\beq
\langle H_Q|\hat T|H_Q\rangle =
-\frac{1}{k^2}\sum_{n=0}^\infty a_n
\left( -\frac{2\La kv}{k^2+i\epsilon}\right)^n ;
\label{30}
\eeq
all trace terms are omitted here since they obviously convolute $k$'s
with each other, and each $k^2$ in the numerator brings in a relative
suppression $\propto \La /m_Q$.

A consequence that immediately stems from this result is the
existence of a ``scaling'' variable. The line shape is determined by a
function depending on the $\phi$ energy only in the ``scaling''
combination
\beq
x=-\frac{k^2}{2\La kv} =
\frac{2}{\La} (E- \frac{m_Q}{2}) .
\label{31}
\eeq
In the physical domain a formal interval of variation of the variable
$x$ is $(-m_Q/\La ) \leq x \leq1$, however to establish the
duality relations one as usual has to consider eq. (\ref{30}) in the
complex
plane.
Of course, it is clear that
on the
negative side the distribution dies off steeply at $x < -x_0$ where
$x_0$ is a positive number of order one. It is worth emphasizing that
$x_0$ needs not be one, and its value is set dynamically, in
distinction with deep inelastic scattering where the limits of
variation of the Bjorken variable are established kinematically,
$0\leq x_{Bj}\leq 1$.

We now can continue our parallel with deep inelastic scattering
further.  Let us show that the coefficients $a_n$ are nothing else but
the moments of a universal distribution function $F(x)$, the function
determining the line shape. Namely,
\beq
a_n=\int_{-\infty}^1 dx x^n F(x) ,\,\,\,  n =0,1,... .
\label{32}
\eeq
We will see in a moment that on general grounds $F(x)$ must be
obviously
positive and indeed its support lies below $1$ as is stated by eq.
(\ref{32});
moreover
$F$ must fall off exponentially towards negative infinity, so that
practically the integration interval is limited by $-x_0$ from below.

We need to emphasize at this point that the distribution function
$F(x)$
introduced so far describes the ``primordial'' heavy quark motion
inside the hadron and does not contain the effect of the hard gluon
emission in the course of the $Q\rightarrow q+\phi$ transition. This
emission can create a light hadronic system  of arbitrary invariant
mass in the final state, which would correspond to a long $
\alpha_s/(1-x)$ tail in the line shape at the negative values of
$x$.
This perturbative tail should be considered separately, of course.

Substituting eqs. (\ref{32}) and (\ref{30}) in eq. (\ref{9})
we arrive at
$$
\frac{d\Gamma}{dE}= -\frac{4}{\pi}\Gamma_0\frac{m_Q E}{M_{H_Q}}
 \Im \int dy F(y) \frac{1}{k^2 +2y\La kv + i\epsilon}
$$
\beq
=\frac{2}{\La}\Gamma_0 F(x)\left( 1+
{\cal O}(\frac{\La}{m_Q})\right)
\label{33}
\eeq
where $x$ is defined in eq. (\ref{31}).  We see that the primordial
heavy quark motion does indeed smear the parton delta function in
$\Gamma_0^{-1}d\Gamma /dE$: instead of the delta function we
have a peak centered at $m_Q/2$ whose width is
$\sim\La$ and height is $\sim 1/\La$ (see Figs.
1, 2). Notice that the normalization condition
\beq
\int dE\frac{d\Gamma}{dE} =\Gamma_0
\label{34}
\eeq
taking place in our approximation is automatically satisfied since
\beq
\int dxF(x) =a_0 =\langle H_Q|\bar Q Q|H_Q\rangle =1 .
\label{35}
\eeq
For a few higher moments it is not difficult to get
\beq
\int dx x F(x) =a_1 =0,
\label{36}
\eeq
\beq
\int dxx^2F(x) =a_2 =\frac{1}{3\La^2}
\langle H_Q|\bar Q{\vec\pi}^2 Q|H_Q\rangle ,
\label{37}
\eeq
\beq
\int dxx^3F(x) =a_3 =-\frac{1}{6\La^3}
\langle H_Q|\bar Q(D_\mu G_{\mu 0}) Q|H_Q\rangle .
\label{38}
\eeq
The gluon operator $D_\mu G_{\mu 0}$ in eq. (\ref{38}) emerges
from the commutator $[\pi_\mu [\pi_\mu \pi_0 ]]$ and  reduces to
the light quark current by virtue of the equations of motion. The
occurrence of this operator, not related to powers of the {\em spatial}
momentum $\vec\pi$, implies that the distribution function
$F(x)$
can {\em not} be interpreted as a non-relativistic heavy quark wave
function in the momentum space.

Summarizing, we presented the procedure of introduction of the
universal distribution function describing the heavy quark motion, so
far in a simplified environment of our toy example. The procedure is
quite similar to what is usually done in deep inelastic scattering,
with certain peculiarities, though. The non-perturbative parameter
$\La$ plays here a role analogous to the nucleon mass in
deep
inelastic scattering (although we will see that it is literally so only as
long as the
effect of perturbative corrections is neglected). As in the latter case
the ``primordial'' function
$F(x)$ is not calculable in perturbative QCD. At the same time, the
renormalization of $F(x)$ due to gluon exchanges is in principle
calculable
although at this point the straightforward similarity with DIS does
not
hold.

Concluding this section it is instructive to make a few important
remarks of a general nature. First, in spite of the non-relativistic
nature of the heavy quark motion inside $H_Q$ the genuine
distribution function $F(x)$ is in no way equivalent to the
non-relativistic (momentum-space) wave function of the heavy
quark; rather it is given (for light final quarks) by the distribution
over
the light cone combination of momentum components.
The distinction, as we discussed above, see eq. (\ref{38}),
becomes apparent already for the third
moment. In other words, all non-relativistic models (say, the
AC$^2$M$^2$
model
\cite{ACM} and other models of this type) are doomed to be
incompatible with the genuine QCD-based
picture of the heavy quark motion, although numerically in some
parts of the spectrum the models can come close to what is expected
from QCD. (This happens, for instance, with the {\em shape} of the
spectrum in the inclusive semi-leptonic decays, see ref. \cite{5}). One
should not expect from these models to reproduce all and every
details of the QCD-based predictions \cite{randal1}.

Second, the theory developed is not applicable in the tails of the
distribution. One of the tails, at negative values of $x$, has been
already mentioned. The hard gluon emission in the transition
$Q\rightarrow q+\phi$ populates the domain $x<-x_0$
in the spectrum at the level $\alpha_s/(1-x)$ which is
much higher than the primordial heavy quark motion. (For more
details see Sect. 5).
Thus, the negative  tail of the ``primordial'' $F(x)$ is simply buried in
the intensive perturbative  background of $Q\rightarrow q+ {\rm
gluon} +\phi$.  Another change of regimes happens at positive values
of $x$, very close to the end-point, $E=M_{H_Q}/2$. When
$E-(M_{H_Q}/2)\sim \La^2/m_Q$ the invariant mass of the
final hadronic state produced in the transition is of order
$\Lambda^2$, not $\Lambda m_Q$. Formally the operator expansion
for the
transition operator $\hat{T}$ we used is not applicable in the end
point
region if one aims at such high energy resolution because the
corresponding
operator enters then in the complex plane too close to the singular
point.
Physically it means that this small
end-point interval of the spectrum belongs to the resonance domain,
and the inclusive approach we have used for developing our
theory is inapplicable. All probabilities in this small end-point
interval are definitely suppressed by additional inverse powers of
$m_Q$. For very large $m_Q$ one could try \cite{BD} to use the
form-factor
type approaches \cite{chernyak} for a quantitative description that
accounts
for the exchanges of semi-hard gluons. For
practical purposes of treating  the $b$ quark decays these
approaches do not seem to be useful, however, since the $b$ quark is
presumably too light to justify the use of the asymptotic form-factor
methods.

\section{Real QCD: $B\rightarrow \gamma X_s$ and Semi-Leptonic
Decays of Beauty}

In this section the analysis carried out above in the toy example will
be extended to the actual processes one encounters with in the
genuine QCD,  namely the radiative transitions of the type
$B\rightarrow \gamma X_s$ and the semi-leptonic inclusive decays
of the type $B\rightarrow  l{\bar\nu}_l X_u$. For simplicity in this
section we  drop our generic notations $Q$ and $H_Q$ and  speak
about $b$ quarks and $B$ mesons. The results will be equally
applicable to any heavy quark $Q$ and any $Q$ containing hadron,
meson or baryon. In those cases when the spin of the decaying heavy
hadron is  non-zero the averaging over spin is implied.

We start from a purely kinematic question of introduction of the
structure functions. Quite obviously, the transition operator $\hat T$
now becomes a Lorentz tensor,
\beq
{\hat T}_{\mu\nu} = i\int d^4x {\rm e}^{-iqx}
T\{ j_\mu^\dagger (x) j_\nu (0)\}
\label{39}
\eeq
where the current $j_\mu$ depends on the process considered.
For the semi-leptonic decays
\beq
j_\mu =\bar u \gamma_\mu (1-\gamma_5) b
\label{40}
\eeq
while for the radiative transitions (neglecting the $s$ quark mass)
\beq
j_\mu = i\partial_\nu [\bar s (1+\gamma_5)\sigma_{\mu\nu} b].
\label{41}
\eeq
Averaging ${\hat T}_{\mu\nu}$ over the unpolarized beauty state
one gets the forward scattering amplitude $h_{\mu\nu}$ \cite{2}
$$
h_{\mu\nu} = \frac{1}{2M_B}
\langle B|{\hat T}_{\mu\nu}|B\rangle =
$$
\beq
 -h_1\,g_{\mu \nu}  + h_2\,v_\mu\,v_\nu
     -i\, h_3\,\epsilon_{\mu\nu\alpha\beta}\,v^\alpha q^\beta +
   h_4\,q_\mu\,q_\nu + h_5\,(q_\nu\,v_\mu+q_\mu\,v_\nu),
\label{42}
\eeq
where the hadronic functions $h_i$ depend on two variables, $qv$
and $q^2$ and $\epsilon_{0123}=-1$. The measurable inclusive rates
are expressible through the imaginary parts of $h_i$,
\beq
w_i = 2\, \Im\,h_i .
\label{43}
\eeq
In full analogy with deep inelastic scattering we will refer to
$w_i(qv, q^2)$ as to the structure functions. In the most general case
there are five structure functions (not six, as indicated in \cite{2}).
They are, clearly, different for the semi-leptonic and the radiative
transitions. Moreover, the generic decomposition of eq. (\ref{42})
significantly simplifies in the case of the radiative transitions
because of the transversality of the current (\ref{41}) and due to the
fact that here $q^2=0$. Concretely,
\beq
\frac{1}{i} \mbox{disc} (h_{\mu\nu})_{B \rightarrow \gamma X_s}
= g_1 (-g_{\mu\nu}(qv) + q_\mu v_\nu +q_\nu v_\mu ) (qv)
-ig_2\epsilon_{\mu\nu\gamma\delta}v^\gamma q^\delta (qv)
\label{44}
\eeq
where $g_1$ and $g_2$ are functions of $qv$.

To complete the kinematical analysis let us give the expressions for
the differential rates (inclusive with respect to the final hadronic
states). In the semi-leptonic decays
$$
{d^3\Gamma(B\rightarrow l{\bar\nu}_lX_u) \over dE_l\,dq^2dq_0}=
$$
\begin{equation}
|V_{ub}|^2 {G_F^2\over 64\,\pi^4}\,
[2 q^2 w_1+ [4\,E_l(q_0-E_l)-q^2]\,w_2+2\,q^2(2\,E_l-q_0)\,w_3\,].
\label{45}
\end{equation}
while in the radiative transitions
\beq
\frac{d\Gamma}{dq_0}(B\rightarrow\gamma X_s)
=\frac{\lambda^2}{2\pi^2}q_0^3 \, g_1 .
\label{46}
\eeq
Here $q_0 = qv$, $V_{ub}$ is the CKM matrix element, $G_F$ is
the Fermi constant and finally the coupling constant $\lambda$
parametrizes the strength of the $b\rightarrow s\gamma$
transition \footnote{The strange quark mass is neglected.},
\beq
{\cal L}\vert_{b\rightarrow s\gamma} =
\frac{i}{2}\lambda F_{\mu\nu}\bar s (1-
\gamma_5)\sigma_{\mu\nu} b ,  \; \; \;
\Gamma_0 (b\rightarrow s\gamma) = \frac{ \lambda^2}{4\pi}
m_b^3.
\label{47}
\eeq

It is worth noting that $w_4$ and $w_5$ do not enter in the
semi-leptonic rate due to the fact that we neglected the charged
lepton
mass. Moreover, in the spectrum of the radiative transitions the
function $g_2$ drops out; to measure $g_2$ it is necessary to detect
the difference in the number of the left-handed and right-handed
circular polarized $\gamma$ quanta,
\beq
{N_+ - N_- \over N_+ + N_-} = {g_2 \over g_1} .
\label{48}
\eeq

We proceed further in close analogy to our toy example. There are
some distinctions, however, which deserve mentioning right now.
First, we deal now with several structure functions:
$w_i$  ($i=1,...,5$) in the semi-leptonic decays and $g_i$ ($i=1,2$) in
the radiative transition. All of them will be expressed in terms one
and the same distribution function $F(x)$ of the heavy quark $b$
in the $B$ meson.  Thus, $F(x)$ is a universal distribution in the same
sense as the light quark distribution functions in deep inelastic
scattering.

The second point, also quite similar to deep inelastic scattering, is the
scaling feature. Namely, the variable $x$ has the form
\beq
x=-\frac{k^2}{2\La kv} =-\frac{m_Q^2+q^2-2m_Q(qv)}{2\La (m_Q-
qv)} ,
\label{49}
\eeq
$$
k_\mu = m_Qv_\mu - q_\mu
$$
and all structure functions $w_i$ which, in general, depend on two
variables, $k^2$ and $kv$,  are actually expressible in terms of the
function $F(x)$ depending on the single variable, the ratio $k^2/kv$.
Of course, in the case of the radiative transition two independent
variables kinematically degenerate into one.

The scaling under discussion is an exact analog of the Bjorken scaling
in deep inelastic scattering and, likewise, is broken by perturbative
loop corrections (which are calculable) and by non-perturbative
power corrections (higher twists).

It is already quite clear from Sect. 2 that our formalism is based on
the
consideration of the transition operator given, in the case at hand, by
eq. (\ref{39}). In the tree approximation corresponding to the graph
of Fig. 3 we substitute the current (\ref{40}) generating the weak
semi-leptonic decay in eq. (\ref{39}) and get
\beq
{\hat T}_{\mu\nu} =-2\int d^4x
(x|\bar b\gamma_\mu\frac{1}{\not\!\! k +\not\!\!\pi}
\gamma_\nu (1-\gamma_5) b|0)
\label{50}
\eeq
where the Schwinger technique is used again. After simple algebra
the expression (\ref{50}) is transformed into
\beq
{\hat T}_{\mu\nu} =-2\int d^4x
(x|\bar b\gamma_\mu ({\not\!\! k +\not\!\!\pi})
\gamma_\nu (1-\gamma_5)
\frac{1}{k^2 +2k\pi +\pi^2 +(i/2)\sigma_{\mu\nu}G_{\mu\nu}}
b|0).
\label{51}
\eeq
In the leading approximation one can neglect $\not\!\!\pi$ compared
to $\not\!\! k$ in the numerator and $\pi^2
+(i/2)\sigma_{\mu\nu}G_{\mu\nu}$ in the denominator. Moreover,
the product of three $\gamma$ matrices reduces to
$\gamma_\alpha$ and $\gamma_\alpha\gamma_5$. The operators
with $\gamma_\alpha\gamma_5$ drop out upon averaging over the
unpolarized hadronic state, while $\gamma_\alpha$ can be
substituted by
$v_\alpha$.  Assembling all these elements together we arrive at
$$
h_{\mu\nu} =\frac{1}{2M_B}\langle B|{\hat T}_{\mu\nu}|B\rangle =
[-g_{\mu\nu}(kv)+k_\mu v_\nu +k_\nu v_\mu -
i\epsilon_{\mu\nu\alpha\rho}k^\alpha v^\rho ]\times
$$
\beq
\frac{1}{2M_B}\langle B|\bar b \left( -\frac{2}{k^2}\right)
\sum_{n=0}^\infty \left( -\frac{2k\pi}{k^2}\right)^n b|B\rangle .
\label{52}
\eeq
As explained in Sect. 2 only the leading twist operators will
contribute in our approximation. Their matrix elements are defined
as
\beq
\frac{1}{2M_B}\langle B|\;{\cal S}\;\,\bar b
\pi_{\mu_1}...\pi_{\mu_n}b-\;
\mbox{traces }|B\rangle =
 a_n{\La}^n (v_{\mu_1}...v_{\mu_n} -\; \mbox{traces}).
\label{53}
\eeq
where ${\cal S}$ is the symmetrization symbol and $\La =
M_B-m_b$.

Next, we introduce the distribution  function $F(x)$ {\em via} its
moments,
\beq
a_n=\int dx x^n F(x) ,\,\,\,  n =0,1,...
\label{54}
\eeq
The lower limit of integration is to be specified dynamically, the
upper limit is 1 (cf. eq. (\ref{32})).
For the first three moments of $F(x)$ we have expressions perfectly
analogous to those quoted in eqs. (\ref{36}) -- (\ref{38}),
\beq
\int dx x F(x) =a_1 =0,
\label{36a}
\eeq
\beq
\int dxx^2F(x) =a_2 =\frac{1}{3\La^2}(2M_{B})^{-1}
\langle B|\bar b{\vec\pi}^2 b|B\rangle ,
\label{37a}
\eeq
\beq
\int dxx^3F(x) =a_3 =-\frac{1}{6\La^3}(2M_{B})^{-1}
\langle B|\bar b(D_\mu G_{\mu 0}) b|B\rangle .
\label{38a}
\eeq

Calculation of the discontinuity of
$h_{\mu\nu}$ is straightforward and leads to the following structure
functions:
$$
w_1 =\frac{2\pi}{\La} F(x) ,
$$
\beq
w_2=\frac{2m_b}{kv}w_1,\,\,  w_3=\frac{1}{kv}w_1,\,\,
w_4=0,\,\, w_5=-\frac{1}{kv}w_1.
\label{55}
\eeq
Notice that all these expressions are transparent generalizations of
the formulae referring to the free quark decay. The latter case, the
free quark decay, is recovered by the substitution
$$
F(x)\rightarrow\delta (x) .
$$

Absolutely similar procedure yields the following results for the
functions $g_i$ in the radiative decay:
\beq
g_1=g_2=\frac{8\pi}{\La} F(x)
\label{56}
\eeq
where now
\beq
x=\frac{2}{\La} (q_0- \frac{m_b}{2})
\label{57}
\eeq
(cf. eq. (\ref{31})).

After the structure functions are found mere substitutions yield
the inclusive distributions sought for,
\beq
\frac{d\Gamma (B\rightarrow\gamma X_s)}{dq_0}
=\Gamma_0(b\rightarrow s\gamma )\frac{2}{\La}F(x) ,
\label{58}
\eeq
\beq
\frac{d\Gamma (B\rightarrow l \bar{\nu}_l X_u)}{dE_l dq^2 dq_0}=
\Gamma_0 (b\rightarrow l \bar{\nu}_l\, u) \frac{2}{\La}
F(x)
\frac{12(q_0 - E_l)(2m_bE_l - q^2)}{(m_b - q_0)}
\label{59}
\eeq
where
\beq
\Gamma_0 (b\rightarrow l \bar{\nu}_l\, u)= |V_{ub}|^2
\frac{G_{F}^2 m_b^5}{192\pi^3}
\label{60}
\eeq
and the argument $x$ of $F(x)$ is defined by eq. (\ref{57}) for
$B \rightarrow \gamma X_s$ and by eq. (\ref{49}) for
$B\rightarrow l \bar{\nu}_l\,X_u$.

Notice the scaling feature of
the differential distribution (\ref{59}), the structure functions
$w_i (q_0, q^2)$ of two variables are expressed {\it via} a function of
one variable $x=(2m_b q_0 - m_b^2 -q^2)/(m_b - q_0)$. This
statement is a full analog of the Bjorken scaling in DIS. The parallel
goes further, of course. Like in DIS, perturbative gluons will lead to a
breaking of the scaling law. The perturbative violations of scaling are
calculable; some aspects of the radiative gluon corrections will be
discussed in the next section.

We pause here to warn the reader: one should be cautious in
applying eq. (\ref{59}) in the entire allowed kinematic domain. It is
not valid in certain boundary subdomain.  For instance, for the
maximal value of the $q^2$,
$$
q^2_{max} = 2M_BE_l ,
$$
the differential rate turns out to be negative.  This happens because
our approximation fails in the domain where the invariant mass
squared of the hadronic state produced is of order
${\La}^2$,
not $\La m_b$.  The error due to the inaccuracy of eq.
(\ref{59}) in the boundary subdomain shows up  in the integral
quantities, like, say, the total decay rate, etc.,  at the level of
${\La}^2/m_b^2$. In the present  paper we do not discuss
these
quadratic effects.

Integrating eq. (\ref{59}) we get the result for the lepton energy
spectrum  in the form
\beq
\frac{d\Gamma (B\rightarrow l \bar{\nu}_l X_u)}{dE_l}=
\Gamma_0 \, \int_{(2E_l -m_b)/\La}^1 d x \, F (x) \,
\frac{16E_l^2}{m_b^3} (3m_b - 4 E_l + \La x).
\label{61}
\eeq
The term $\La x$ in the brackets is of a kinematical origin.
Corrections to the spectrum of the same order,
${\cal O}(\La /m_b)$, are generated also dynamically,
{\em via} higher twist operators in OPE which appear if one accounts
for the terms $\not\!\!\pi$ and $\pi^2 +(i/2)\sigma G$ in eq.
(\ref{51}). In the above analysis these terms were neglected in the
expansion. The higher twist effects can be treated essentially in the
same manner it is usually done in DIS, see e.g. \cite{shuryak}.

If the energy $E_l$ is not too close to its upper limit
eq. (\ref{61}) reduces to
the  spectrum of the free quark decay. Indeed, when
$(m_b - 2E_l) >> \La$, the  lower limit of integration in
eq. (\ref{61}) becomes a large (negative) number, and the integrals
over $x$ can be done explicitly ( see e.g. eq.  (\ref{36a}))
$$
\int_{-\infty}^1 dx F(x) =1,\,\,\,  \int_{-\infty}^1 dx x
F(x)=0 .
$$
Of most interest is the range of energies in the {\em window},
\beq
0< 2E_l -m_b <\La
\label{61a}
\eeq
and nearby. For such energies the lower limit in the
 integral (\ref{61}) becomes of order  1, and  the end point shape of
the
lepton spectrum is expressed {\it via} integrals over the distribution
function
$F(x)$. Although this function is unknown, it is the same one that
determines the  inclusive radiative transitions.

\section{Including the Mass of the Final Quark}

Here we address the question  what happens
if the final quark mass is non-vanishing. In general, the primordial
motion of the initial heavy quark will now manifest itself {\em via}
{\em different} distribution functions. This shows again that
non-relativistic models of the AC$^2$M$^2$ type are too gross to
reproduce all subtle features stemming from QCD. We plan to discuss
the issue in more detail elsewhere. Here, just to illustrate our point,
we will consider the limit of the non-relativistic final quark and
again resort to the toy example of Sect. 2. The quark spins play no
role
in this limit, of course, and can be ignored.

If the initial quark mass is $m_Q$ and that of the final quark is
$m_q$ we assume that
\beq
\Delta m = m_Q -m_q << m_Q .
\label{B}
\eeq
Of course, the validity of the inclusive description we consistently
use requires simultaneously that
\beq
\Delta m >>\Lambda .
\label{C}
\eeq
This is the so-called small velocity (SV) limit \cite{SV}. In this limit
the velocity $\vec v$  of the final quark in the transition $Q\ra q
\phi$
is
$$
|\vec v| = \Delta m /m_q .
$$
Notice that $\Delta m$ coincides with $M_{H_Q}-M_{H_q}$, and the
physical velocity of the final hadron is the same as above.

In the SV limit the $\phi$ spectrum is peculiar. At the free quark
level
it consists of a single monochromatic line, similar to eq. (\ref{6}),
$$
\frac{d\Gamma}{dE} =\Gamma_0\delta (E-E_0),\,\,\,
E_0 = \Delta m = M_{H_Q}-M_{H_q}.
$$
If we  trace only terms ${\cal O}(v^0)$ the physical spectrum is
exactly the same -- the delta function peak residing at the same
place -- and
the inclusive probability is completely saturated by one heavy
meson in
the final state \cite{SV}. Notice that there is no `window' -- no shift
is present between the
maximal allowed energies of $\phi$ at the quark and the hadron
levels in the case at hand.

Modifications of this perfect quark-hadron duality
start at the level of ${\cal O}(v^2)$ \cite{bjorken2}.  If ${\cal O}(v^2)$
effects are considered
 the height of the elastic peak is changed, and a comb of
inelastic peaks appears, the height of the latter being proportional to
$v^2$. This comb will lie at $E<E_0$ and will be stretched over an
interval of $\phi$ energies of order $\Lambda$. (As was mentioned
above, $\La$ is not a relevant parameter in this problem;
we can continue to use it, however, just as a typical hadronic scale.
One could certainly choose another definition of the typical hadronic
scale.) The integral over the inelastic peaks must compensate the
distortion of the elastic one -- a  variant of the Bjorken sum rule
\cite{bjorken2} taking place in our toy example. What we would like
to do is to relate the inelastic spectrum (the comb of resonances in
the final state) to a new distribution function connecting this
spectrum
to the motion of the initial heavy quark inside $H_Q$.

To this end we again turn to consideration of the transition operator
(\ref{8}). Our analysis is changed in a minimal way --
eq. (\ref{27}) is substituted by
\beq
\hat T = \frac{1}{m_q^2-k^2}\sum_{n=0}^\infty
\bar Q \left( \frac{2m_Q\pi_0 +\pi^2- 2q\pi }{m_q^2-k^2}\right)^n Q .
\label{S1}
\eeq
where $q$ is the momentum of $\phi$. The SV limit means that
$q$ is to be treated as a small parameter, and we will expand in $q$
keeping the terms up to the second order. In the zero-th order in $q$
the equation of motion
$$
\left( \pi_0 +\frac{\pi^2}{2m_Q} \right) Q=0
$$
immediately tells us that the corrections to the free quark result are
absent (up to an overall normalization),
$$
\langle H_Q|{\hat T}^{(0)}|H_Q\rangle  = \frac{1}{m_q^2-k^2}
\langle H_Q|\bar Q Q|H_Q\rangle =
$$
\beq
\frac{1}{m_q^2-k^2} \, \frac{M_{H_Q}}{m_Q} \left( 1+ {\cal O}
(\La^2/m_Q^2)\right) .
\label{S2}
\eeq

In the first and second order in $q$ we get
\beq
\langle H_Q|{\hat T}^{(1)}|H_Q\rangle =-\frac{q_0}{m_Q}\,
\frac{\langle{\vec\pi}^2\rangle}{(m_q^2-k^2)^2} ,
\label{S3}
\eeq
\beq
\langle H_Q|{\hat T}^{(2)}|H_Q\rangle = \frac{4}{3}{\vec q}^2
\frac{1}{(m_q^2-k^2)^3}
\sum_{n=0}^\infty \left( \frac{2m_Q}{m_q^2-k^2}\right)^n
\langle \pi_i\pi_0^n\pi_i\rangle .
\label{S4}
\eeq
To compress our formulae here and below we resort to a short-hand
notation, for instance,
$$
\langle \pi_i\pi_0^n\pi_i\rangle\equiv
\langle  H_Q|\bar Q\pi_i\pi_0^n\pi_i Q|H_Q\rangle .
$$

What follows just parallels all steps leading to the introduction of the
function $F(x)$. We introduce now a new distribution function $G(x)$
whose moments $b_n$ are realted to the matrix elements of the
operators in eq. (\ref{S4}),
\beq
\langle \pi_i\pi_0^n\pi_i\rangle = b_n\La^{n+2} ,
\label{S5}
\eeq
\beq
\int dx x^n G(x) = b_n .
\label{S6}
\eeq
It is evident that $G(x)$ can be called a `temporal' distribution
function, unlike $F(x)$ which may be called the `light-cone'
distribution
function.

Notice that $b_0 =3a_2$ and is related to $\langle
{\vec\pi}^2\rangle$
while $b_1 =3a_3$ and is related to $\langle (D_iE_i)\rangle$
where $E_i$ is the chromoelectric field. For $b_n$ with $n\geq 2$
the operators $\pi_i\pi_0^n\pi_i$ are expressible in terms of the
chromoelectric field and its time derivatives,
\beq
\langle \pi_i\pi_0^n\pi_i\rangle = i^{n-2}
\langle E_i (D_0^n E_i)\rangle .
\label{S7}
\eeq

Substituting eqs. (\ref{S5}), (\ref{S6}) in eqs. (\ref{S2}) --
(\ref{S4}) we obtain for the imaginary part of the transition operator,
$$
{\rm Im} \, (2M_{H_Q})^{-1}\langle H_Q|{\hat T}|H_Q\rangle =
\frac{\pi}{4m_Q^2\La}\left[ \delta (x)
\left( 1-\frac{1}{3}\frac{{\vec q}^2}{m_Q^2}\int dy y^{-2} G(y) \right)
+
\right.
$$
$$
\delta ' (x)\left( \frac{q_0\langle {\vec\pi}^2\rangle}{2m_Q^2\La}
+\frac{1}{3}\frac{{\vec q}^2}{m_Q^2}\int dy y^{-1} G(y) \right) +
$$
\beq
\left.
\frac{1}{3}\frac{{\vec q}^2}{m_Q^2} x^{-2}G(x) \right]
,
\label{S8}
\eeq
where the variable $x$ is
$$
x=\frac{m_q^2-k^2}{2m_Q\La} = \frac{E-E_0}{\La},\,\,\,
E_0 =\frac{m_Q^2-m_q^2}{2m_Q} .
$$
The first line in eq. (\ref{S8}) describes the elastic peak renormalized
by ${\vec q}^2$ corrections. The second line is a small shift in the
position of
the elastic peak. Finally, the third line represents the inelastic
contribution which, as we see, is proportional to ${\vec q}^2$ in full
accord
with the general expectations \cite{bjorken2}.

If we now combine this imaginary part with the general expression
(\ref{9}) we arrive at the physical spectrum $d\Gamma /dE$ in the
$H_Q\ra
\phi X_q$ decay. In principle, eq. (\ref{S8}) has a wider range of
validity since it is applicable for $q^2\neq 0$ as well. In the case
$q^2=0$ one must substitute
$$
q_0 =|\vec q | = E = E_0 +\La x .
$$

Different terms in eq.  (\ref{S8}) are of the different order in
$\La /m_Q$. In the leading order in $\La/m_Q$ the second line must be
omitted and ${\vec q}^2$ in the first and the third lines must be replaced
by $E_0^2$. Moreover, the overall factor $E$ relating eq.
(\ref{S8}) to the spectrum (\ref{9}) must be replaced by $E_0$ as
well.

If one calculates now the total decay width  $\Gamma$ by
integrating over the $\phi$ energy the terms proportional to
${\vec q}^2$ in the first and the third lines cancel each other and the
total width obtained in this way coincides with the free quark one.
The
suppression of the elastic peak is exactly compensated by the
integral over the inelastic part of the spectrum. Non-renormalization
of the total width in the order ${\cal O}(v^2)$ is nothing else than the
 Bjorken sum rule \cite{bjorken2}.

This assertion of the non-renormalization of the total width should,
of course, survive inclusion of the $\La /m_Q$ corrections simply
because these corrections, as we know \cite{2,5}, are absent in the
total width. It is instructive to trace how this cancellation is
arranged.

It is a rather straightforward exercise to keep corrections of the first
order in $\La /m_Q$ in the spectrum, and then to check that the
cancellation persists, and the Bjorken sum rule still holds.  Now we
can not use the substitution $E\ra E_0$, of course; instead
we must use $E=E_0 +\La x$, and the second line in eq. (\ref{S8})
must also be taken into account. One should also keep in mind that
the function $G(x)$ itself was defined above only in the leading
order. We can generalize the definition, however, to include
corrections $\La /m_Q$ in a corrected definition of $G(x)$.  What is
important is that the same new (corrected) function will appear in all
lines in eq. (\ref{S8}), which will guarantee the desired cancellation
and the validity of the non-renormalization theorem
\cite{SV,bjorken2}.

Let us now discuss in brief the general case of the arbitrary ratio
$$
\gamma =\frac{m_q}{m_Q} .
$$
The physical boundary between what can be called here the light
final quark and the heavy one lies at
\beq
m_q^2 \sim \La m_Q .
\label{S20}
\eeq
If $m_q << (\La m_Q)^{1/2}$ then the effects due to $m_q$ are small.
On the other hand, to the right of the boundary (\ref{S20}) these
effects become of order one, and the question arises as to how one
can generalize the analysis above.

The generic distribution function for the arbitrary mass ratio
$\gamma$  is introduced through the matrix elements of the
operators
\beq
\frac{1}{2M_{H_Q}\La^n}
\langle H_Q|\bar Q (\nu\pi )^n Q|H_Q\rangle =
\int dx x^n F_\nu (x) ,
\label{S21}
\eeq
where the four-vector $\nu$ is defined as
$$
\nu_\mu =\frac{2k_\mu}{m_Q},\,\,\, \nu^2 =4\gamma^2 .
$$
It is not difficult to check that for $\gamma =0$ we return back to
the definition of the `light-cone' distribution function, and $F_\nu$
coinsides with $F$, eqs. (\ref{53}), (\ref{54}). In the opposite limit
$\gamma\ra 1$ the function $F_\nu$ is proportional to
$(1-\gamma )^2 G$ where $G$ is the `temporal' distribution function.

Thus, although the heavy quark distribution functions introduced are
universal for each given value of $\gamma$, for different $\gamma$'s
we have to deal with different distributions.  This is another
manifestation of the fact that the naive models of the AC$^2$M$^2$
type are
intrinsically incompatible with QCD.

We pause here to make several remarks. Notice that at
$\gamma\neq 0$ and $q^2\neq 0$ the scaling law (the dependence
on a single variable, eq. (\ref{49})) does not hold any more. We do
not know what replaces it.

Our next observation concerns the size of the window. It is quite
obvious that the window shrinks as $\gamma$ approaches 1, as
$(1-\gamma )^2\La$. In other words,
$$
F_\nu (x) =0\,\,\, {\rm at}\,\,\, x>\frac{1}{2} (1-\gamma )^2 \La .
$$
This property implies a specific behavior of the matrix elements
(\ref{S21}) at large $n$.  It is clear that if $\gamma$ is close to 1,
when the window is essentially absent, the introduction of the
parameter $\La$ is purely artificial. Any dimensionful parameter of
the typical hadronic size could have been used for parametrization of
the matrix elements (\ref{S21}).

Needless to say that the analysis carried out above is trivially
generalizable to the  case of real QCD. The strange (and the more so
$u$ and $d$) quarks produced in the $b$ decays can be treated as
light (massless). As for the $c$ quark, it lies close to the boundary
(\ref{S20}), $m_c^2\sim $2 GeV$^2$ and $\La m_b\sim$ 2 GeV$^2$.
Thus, formally it does not belong to either of the limits. Still
arguments exist showing that the non-relativistic description of the
$c$ quark in the inclusive $b$ decays is not so far from reality.

\section{Radiative Corrections}

The above discussion of the end-point behavior of the spectra refers
only to purely nonperturbative effects. In reality the perturbative
corrections
are also present. As we know from  experience with DIS the
perturbative
corrections are not small because of the presence of large logarithms
and lead to violation of the
scaling behavior. We need an analog of the Lipatov-Altarelli-Parisi
evolution kernel to account for the  perturbative effects. The
primordial
distribution will be convoluted with this  evolution kernel.

Let us first sketch a qualitative picture using the example of
$B\rightarrow \gamma X_s$.  In the free quark approximation the
photon spectrum is a monochromatic line (the delta function in eq.
(\ref{6})). The primordial spread of the heavy quark momentum
substitutes the monochromatic line by a finite width line. We account
for the primordial spread by convoluting the primordial momentum
distribution with the delta function kernel (see eq. (\ref{33})). The
width of the line becomes of order $\La$. Outside this
end-point domain whose size is $\sim \La$ the primordial
motion has a negligible impact on the photon spectrum.

 As a  matter of fact, even without the primordial spread, the
perturbative
gluon emission smears the delta function of eq. (\ref{6}), producing
quite a drastic effect both in the end-point domain and,
even more so, outside this domain.  First, a
radiative tail is generated below $E=m_b/2$. If $(m_b/2) -E
\sim (m_b/2)$ the radiative tail is ${\cal O}(\alpha_s)$; it is further
enhanced as one approaches the end-point domain. The
end-point peak in the spectrum looses its delta function shape, but
still a certain singular behavior at $E=m_b/2$ persists. The shapes of
the radiatively distorted peak and the adjacent tail are determined
by the Sudakov exponent. The total probability is redistributed: the
area under the peak in the end-point domain
(which used to be one in appropriate units in the
Born approximation) is now $\sim (\La
/m_b)^{\epsilon_0}$ where $\epsilon_0$ is a positive number, and
the total probability is saturated in the tail. Clearly, the perturbative
gluon corrections do not generate the spectrum in the window
$m_b/2<E<M_B/2$.  To describe the effect of the filling of the
window one has to include the   primordial momentum distribution
of the heavy quarks which is now  to be convoluted with
a new kernel,  created  by the gluon radiative corrections.

The distortion of the delta function (\ref{6}) described above is
physically quite
transparent.  The total area under the distorted
curve is unchanged, however.  The picture must be very familiar to
those readers who remember the treatment of the electromagnetic
radiative corrections in the $J/\psi$ peak in $e^+e^-$ annihilation
\cite{J}.
There, the natural width of the $J/\psi$ meson is also negligibly
small, and the observed shape of the peak is totally determined by
two effects:  the radiative smearing of the original Breit-Wigner
$J/\psi$ peak and the energy spread in the colliding beams. These
two effects are convoluted.

We proceed now to a more quantitative discussion of the issue
concentrating on the example of $B\rightarrow\gamma X_s$. The
conclusions will be of  a more general nature, of course.
The convolution discussed above can be written in the following way:
\beq
\frac{d\Gamma_B(E)}{dE} =  \int dy F(y)
\frac{d\Gamma^{pert}_b(E-(\La/2) y)}{dE}
\label{62a}
\eeq
where $d\Gamma_B(E)/dE$ denotes the observable (physical)
$\gamma$ spectrum while $d\Gamma^{pert}_b/dE$ refers to the
photon spectrum in the $b\rightarrow s+\gamma +$ gluons
transition in perturbation theory. If $d\Gamma^{pert}_b/dE$
is substituted by the delta function distribution (\ref{6}) we recover
the old result (\ref{33}). On the other hand, if the perturbative
smearing produces $d\Gamma^{pert}_b/dE$ that is much broader
(as  function of $x$) than $F(x)$ then the primordial spread $F(x)$
does not affect the physical spectrum $d\Gamma_B(E)/dE$, and the
latter coincides with the quark-gluon one unless one is interested in
small
corrections.

To find what situation is actually realized we need to explicitly
calculate
$d\Gamma^{pert}_b/dE$. As we will see shortly, the shape of the
peak characteristic to  the
primordial momentum spread $F(x)$ is recovered in the end-point
domain. The peak in the physical spectrum near $E=m_b/2$ persists,
although its height is scaled
down significantly.

To begin with let us consider the one gluon emission. If one  limits
oneself
to the double-logarithmic approximation (DLA) the result can be
borrowed from text-books; namely one obtains
\beq
\frac{d\Gamma^{pert}_b}{dE} = -\frac{2m_b}{\pi}\Gamma_0
\Im \left[ \frac{1}{k^2+i\epsilon}\left(
1-\frac{2\alpha_s}{3\pi}\ln^2\frac{m_b^2}{-k^2}\right)\right]
\label{63a}
\eeq
 where $\Gamma_0$ is given by eq. (\ref{47}), $k^2 =
(m_bv-q)^2 = 2m_b(E-(m_b/2))$, and $\alpha_s$ is the gluon
coupling constant. We are mostly interested in the end-point domain
$ (E-(m_b/2))\sim\La$, i.e.
\beq
k^2 \sim m_b\La ;
\label{64a}
\eeq
then the logarithm of $m_b^2/k^2$ can and must be treated as
a large parameter.  Eq. (\ref{63a}) keeps only the double-logarithmic
terms, those with one  log less are omitted.

A subtle point which deserves mentioning in connection with eq.
(\ref{63a}) is an ``enhanced'' singularity at $k^2=0$. For calculating
the imaginary part of eq. (\ref{63a}) {\em per se} at $k^2=0$ one
should, in principle,  regularize the logarithm in the infrared region.
There is
no need in any explicit regularization, however, since the summation
of the  double logarithms  provides us with a natural regularization
-- the point $k^2=0$ is absolutely suppressed after the summation.

Collecting all double logarithms in the standard manner we get the
well-known Sudakov exponent,
\beq
\frac{d\Gamma^{pert}_b}{dE} = \frac{2m_b}{\pi}\Gamma_0
 \Im \left\{ \frac{1}{-k^2}\exp \left[
-\frac{2\alpha_s}{3\pi}\ln^2\frac{m_b^2}{(-k^2)}\right]\right\} .
\label{65a}
\eeq
The fact that the gluon coupling $\alpha_s$ runs is unimportant in
deriving eq. (\ref{65a}) since this expression corresponds to DLA,
while  the running nature of $\alpha_s$ becomes visible only at the
level of subleading logs. With the double $\log$ accuracy
$\alpha_s$
normalized at any point is equally suitable in eq. (\ref{65a}). We can
try to guess, however, what would be the effect
of subleading $\log$s on $\alpha_s$ on physical grounds, without
carrying out a complete and consistent analysis (for a dedicated
discussion see ref. \cite{DOK}). If $k^2$ provides an infrared cut-off
for all transverse momenta in the problem then the largest possible
value of $\alpha_s$ is $\alpha_s(k^2)$. Then
substituting
$\alpha_s \ra\alpha_s(k^2)$ in eq. (\ref{65a}) we only overestimate
the Sudakov suppression, and actually its influence will be milder
than the estimates following below show. We will anyway use the
prescription $\alpha_s \ra\alpha_s(k^2)$ for a simple-minded
modelling of the Sudakov effect. Of course, this aspect can be
improved. Some trivial improvement will be mentioned shortly.

The validity of DLA implies that the condition
$$
\frac{\alpha_s}{\pi}\ln\frac{m_b^2}{-k^2} << 1
$$
must be met.  This condition is numerically satisfied provided that
$k^2$ is chosen in the interval (\ref{64a}), as we will see below.

Next we observe that the Sudakov exponent can be identically
rewritten as a power of $k^2$,
\beq
\exp \left[
-\frac{2\alpha_s(k^2)}{3\pi}\ln^2\frac{m_b^2}{(-k^2)}\right]
=\left(\frac{-k^2}{m_b^2}\right)^\epsilon ,
\label{66a}
\eeq
where the exponent $\epsilon$ is given by the following expression:
$$
\epsilon = \frac{2\alpha_s(k^2)}{3\pi}\ln\frac{m_b^2}{(-k^2)}
$$
provided that the prescription $\alpha_s\ra \alpha_s(k^2)$ is indeed
valid.
If $k^2 =-m_b\La x $ then, obviously,
\beq
\epsilon =\frac{8}{3b}\, \frac{\ln (m_b/\La ) -\ln x}{\ln
(m_b/\La ) +2 \ln (\La /\Lambda ) +\ln x }
\label{67a}
\eeq
where $b$ is the first coefficient in the Gell-Mann-Low function and
$\Lambda$ is the scale parameter of QCD. For $x$ of order one and
large $\ln (m_b/\La )$ this expression approximately
reduces to a constant, to be denoted below by $\epsilon_0$,
\beq
\epsilon_0 =\frac{8}{3b}\approx 0.3 .
\label{68a}
\eeq
The numerical value of $\epsilon_0$ turns out to be rather small, a
fact crucial for our consideration.

Let us parenthetically note the following. If, instead of
$\alpha_s(k^2)$ as an overall factor in the Sudakov exponent, we
kept the running $\alpha_s$ {\em inside} the integrals determining
the double logarithms the only modification of the results (within the
accepted approximations) would be a shift of $\epsilon_0$ towards a
lower value.  This is physically quite transparent since the Sudakov
exponent becomes less suppressive. Furthermore,  subleading terms
given  in  text-books  also work in the same direction: they enhance
the pre-exponential factor tending to compensate in part
the exponential suppression.

Summarizing we can say that the net effect of the Sudakov
suppression reduces to the substitution
\beq
\frac{1}{k^2}\ra \frac{1}{(k^2)^{1-\epsilon_0}(m_b^2)^{\epsilon_0}}.
\label{69a}
\eeq
Correspondingly, the radiatively corrected spectrum takes the form
\beq
\frac{1}{\Gamma_0}\frac{d\Gamma^{pert}_b}{dE}
=
\frac{2}{\pi m_b}\frac{\sin\pi\epsilon_0}{\left(
 1-(2E/m_b)\right)^{1-\epsilon_0}} \theta (\frac{m_b}{2}-
E).
\label{70a}
\eeq
It is not difficult to see that this expression is actually valid as long
as
$|E-(m_b/2)| \ll (m_b/2)$, i.e. we are not necessarily confined to the
end-point domain, eq. (\ref{64a}). Moreover, for qualitative purposes
one can extrapolate  eq. (\ref{70a}) even further down, to $(m_b/2)-
E\sim (m_b/2)$. One sees then that the photon spectrum is
dramatically reshuffled by the gluon radiative corrections: the delta
function at $m_b/2$ is substituted by a much milder peak at this
point, plus the tail which  saturates the probability.

To substantiate the point let us check what  contribution to
$\Gamma$ comes from  the end-point domain by integrating eq.
(\ref{70a}) from $E=m_b/2$ to $E=(m_b/2)-\La$,
\beq
\frac{1}{\Gamma_0}\int_{(m_b/2)-\La}^{m_b/2}
dE\frac{d\Gamma^{pert}_b}{dE} \propto
\left(\frac{\La}{m_b}\right)^{\epsilon_0}.
\label{71a}
\eeq
In other words, the end-point contribution is suppressed as a
(relatively small) power of $m_b^{-1}$.

We pause here to investigate in more detail what happens with the
shape of the {\em physical} spectrum in the end-point domain  when
both
effects -- the perturbative gluon corrections and the primordial
heavy quark motion -- are taken into account. Recall that only the
latter effect populates the spectrum to the right of  the point
$E=m_b/2$. Using the general expression (\ref{62a})  and the double
log result (\ref{70a}) for $d\Gamma^{pert}_b/dE$  we obtain
\beq
\frac{1}{\Gamma_0}\frac{d\Gamma_B}{dx}
=
\left(\frac{\bar\Lambda}{m_b}\right)^{\epsilon_0}
\int_{0}^{1-x} dz F(x+z) \frac{d}{dz}(z^{\epsilon_0})
\label{72a}
\eeq
where $\sin\pi\epsilon_0$ is approximated by $\pi\epsilon_0$ and
$$
x=\frac{2}{\bar\Lambda}\left( E-\frac{m_b}{2}\right) .
$$

Now, the combination of three essential factors --
(i) the fast fall off of the primordial distribution $F(y)$ at large
$y$,
(ii) the specific form of the perturbative kernel (\ref{70a}) and (iii)
the numerical smallness of $\epsilon_0$ -- leads to the fact that the
convolution (\ref{72a}) turns out to be approximately local if $x$ lies
in the end-point domain,
\beq
\frac{1}{\Gamma_0}\frac{d\Gamma_B}{dx}
\sim
\left(\frac{\bar\Lambda}{m_b}\right)^{\epsilon_0}
F(x) .
\label{73a}
\eeq
We hasten to add that the locality in $z$ in eq. (\ref{72a}) is an
approximation valid as long as $\epsilon_0$ can be considered as a
small number. The size of the domain over which $F(y)$ is smeared
by the convolution (\ref{72a}) is of order
\beq
\Delta z_{char} \sim \exp (-c/\epsilon_0)
\label{74a}
\eeq
where the constant $c$ is of order one.  For
$\epsilon_0\sim 0.3$ we have $\Delta z_{char} \lsim 0.1$ which sets
the resolution with which one can extract  the primordial distribution
from the measured photon spectrum in the end-point domain.

A remark is in order here explaining how the estimate (\ref{74a})
is obtained.  Let us take a model for the function $F(x)$ in the range
$0<x<1$ in the form
$$
F(x)=exp(-x/a)
$$
where $a$ parametrizes the width of our model function.
Then the convolution (\ref{72a}) of this function with the perturbative kernel
yields
$${\rm e}^{-x/a}\int_0^{1-x} dz {\rm e}^{-z/a}\frac{d}{dz}
z^{\epsilon_0} \approx {\rm e}^{-x/a} a^{\epsilon_0}\Gamma (1+\epsilon_0)
$$
$$
\approx {\rm e}^{-x/a} [1-\epsilon_0 \ln (1/a)]
$$
provided that $a<<1-x \sim 1$. We also expanded in $\epsilon_0$ and
retain only the term with the potentially large logarithm. If
$\epsilon_0 \ln (1/a) << 1$ the convolution we made reproduces the
original function $F(x)$. At very small values of $a$ where
$\epsilon_0 \ln (1/a) \sim 1$ the Sudakov kernel smearing starts working
in full. In this way we arrive at eq. (\ref{74a}).

Let us try to summarize our findings concerning the role of the
perturbative gluon emission in the transitions like $B\ra\gamma
X_s$.  The  gluon radiative corrections affect strongly the height of
the end-point peak (for the actual $b$ quark this height is
suppressed by
a factor $\sim 2.5$ compared to the result (\ref{58}) containing no
perturbative corrections). The theoretical uncertainty of order ${\cal
O}(\epsilon_0)$ in the overall factor
is due to subleading logarithms which we neglected. At the same
time, the shape of the end-point peak is affected to a much lesser
extent. The primordial
$F(x)$ is smeared, of course, but the width of the smearing is still
much less than unity, see eq. (\ref{74a}). (In the  $B\ra\gamma
X_s$ transition the size of the smearing interval is $\sim$ 30 MeV.)
Since the width of $F(x)$ is about unity
the above smearing does not change the shape drastically.

As for the semi-leptonic decays the triple differential distribution
(\ref{59}) is affected by perturbative gluons in the way very similar
to that just discussed. Correspondingly, the energy spectrum
$d\Gamma (B\rightarrow l \bar{\nu}_l X_u)/dE_l$ is changed: the
integrand in eq. (\ref{61}) acquires an extra factor due to the
radiative corrections. Observe, that the extra integration in
$d\Gamma (B\rightarrow l \bar{\nu}_l X_u)/dE_l$ further smears
$F(x)$; therefore, the energy spectrum is definitely {\em not} the
best  place for determining $F(x)$.  The radiative transitions or the
double differential distribution in the semi-leptonic decays are much
better suited for this purpose.

Below the end-point domain the primordial motion plays no role, and
the physical photon spectrum is fully controlled by perturbative
effects.
The
Sudakov exponent introduces modifications of order unity
parametrically
earlier than nonperturbative effects: the latter appear when $(1-
(2E/m_b))\sim
\Lambda /m_b$ whereas the former come into play already at
$\ln(m_b/(m_b-2E))\sim c
\cdot
(\ln (m_b/\Lambda ))^{1/2}$.

Moreover, one  can account for  the
nonperturbative corrections outside the end-point domain
$\La /m_b \ll (1-(2E/m_b))\ll 1$,
that is just where the perturbative spectrum does not vanish. It can
be done  exactly like in ref.~\cite{5} and the result will be a
{\em regular} series in $1/m_b$ because
the perturbative spectrum is  smooth in this domain.
In particular, no linear corrections to the shape  formally appear.

It is interesting to notice that the power (non-perturbative)
corrections in the domain
$$
\frac{\La}{m_b} \ll \, 1-\frac{2E}{m_b} \, \ll 1
$$
are calculable directly from eq. (\ref{62a}) by expanding
$d\Gamma_b^{pert}/dE$ in $\La y$, without any further
computations. The linear term vanishes due to eq. (\ref{36a}) while
the quadratic term is expressed through eq. (\ref{37a}),
$$
\Delta \frac{d\Gamma_B}{dE}
= \frac{
d^3\Gamma_b^{pert}}{dE^3}\, \frac{1}{24}\, (2M_B)^{-1}
\langle B|\bar Q {\vec\pi}^2 Q|B\rangle .
$$
There are other $1/m_b^2$ corrections, of course, for instance,
the $\sigma G$ term, but they are not enhanced by powers of
$(1-(2E/m_b))^{-1} $.

It should be added that even if
 the amount of the Fermi motion were very small
compared to $\La$ (narrow $F(x)$), the
region close to $M_B/2$ still would be populated due to a totally
different
mechanism. The pick-up mechanism associated with the diagrams
where a
semi-hard gluon
exchanges momentum between the decaying heavy quark and the
light
spectator
produces such final states; it is important that this mechanism does
{\em not} suffer
from
the same Sudakov suppression. This contribution can be analyzed in
a similar
way as it has been done, for example, for the heavy flavor exclusive
decays
into pion or $\rho$ \cite{BD}. It seems definite, however, that for real
beauty
hadrons this mechanism is subdominant.

One more subtle point which deserves mentioning in connection with
our discussion of the gluon radiative corrections. If one goes beyond
the tree approximation a  specific normalization procedure must be
formulated in order to make all operators we work with
well-defined.
This procedure will clearly introduce a normalization point $\mu$.
The operators become $\mu$ dependent through anomalous
dimensions, i.e. standard logarithms of $\mu$ show up. Besides these
standard logs a power dependence on $\mu$ is also present
(see e.g. \cite{novikov}). For instance, the matrix element
$\langle {\vec\pi}^2\rangle$ can contain a piece $\sim
\alpha_s (\mu ) \mu^2$. In practice such terms are difficult to
control, so they represent, rather, a theoretical uncertainty.  For the
OPE machinery to be workable one needs to assume that there exists
a range of $\mu$ where
two contradicting requirements are simultaneously met:
(i) $\alpha_s (\mu )$ is sufficiently small so that QCD perturbation
theory makes sense; (ii) the uncertainty due to $\alpha_s\mu^2$
is much less than the matrix element
$\langle {\vec\pi}^2\rangle$ itself. Experience with the QCD sum
rules teaches us that the requirements can be actually met --
an appropriate choice of $\mu$ is possible (for a review
see \cite{sh}).

In this context the most sensible issue  is the problem of an
unambiguous definition of the parameter
$\La$.  The value of $\La$
{\em per se} does not exist in QCD as a {\em rigorously defined
constant}
(for recent discussion see e.g. \cite{BU4}).  In
reality it is the dominance of the condensates (non-perturbative
matrix elements of relevant operators)
 that enables one to ascribe  more or less definite physical meaning
to this parameter. For our purposes
we can accept the following working definition:
 $\La$ can be understood as the distance between the center of the
end-point peak and the kinematic boundary in $B\ra\gamma X_s$.

The problem with the definition of $\La$ which might seem to be of
a crucial importance,  actually does not affect our analysis. Indeed,
instead of $\La$ we could  use any
fixed hadronic scale, say the QCD scale  parameter $\Lambda$ (the
one
 defining the running coupling constant) in all constructions
presented in Sects. 2, 3 and 4 and in all definitions of the
scaling variables.

\section{Comments on the Literature}

A formal definition of the  heavy quark distribution function inside
the heavy hadron
was first given in Ref. \cite{randal2} in the context of deep inelastic
electroproduction on heavy hadrons. The moments of this function
(which can be called a `light-cone' function) were related to the
matrix elements of the operators (\ref{53}). It is important that the
`light-cone'  distribution function defined by eq. (\ref{53}) appears in
this problem only provided one assumes that $Q^2\gg m_Q^2$.

The prime object of interest for these authors was the heavy quark
fragmentation in heavy hadrons. They did not apply the OPE
machinery discussed above to the heavy hadron decays.

While our definition of $F(x)$ coincides with that of Ref.
\cite{randal2} at the tree level, the perturbative renormalizations
will
be drastically different. The normalization point appropriate
to DIS on heavy mesons considered in \cite{randal2} is $\sim
Q^2\gg m_Q^2$, i.e. academically high. In our case, due to  different
kinematics, the appropriate normalization point is $\sim m_Q\La$.

A similar definition of the `light cone' distribution function was
suggested in \cite{neubert} for the description of the energy
spectrum
near the end-point in the semi-leptonic heavy flavor decays. The
shape function of Ref. \cite{neubert} is related to our $F(x)$ through
an integration.

Our analysis goes further in the following aspects. First, we consider
different processes to underline  the universal nature of the
distribution function $F(x)$. The effect is shown to be much more
pronounced in the radiative decays and in the double differential
distributions in the semi-leptonic decays.

Second, we explain that the `light cone' distribution function,
however important it might be, is by no means a unique
characteristic of the heavy quark motion. Other quantities which are
measurable, at least in principle, can depend on other distribution
functions which reflect the heavy quark motion in a different way.
A clear-cut example is the heavy quark decay into another heavy
quark. If the mass of the quark produced is such that it is
non-relativistic (the SV limit) the spectrum in this transition will
measure
a `temporal' distribution function $G(x)$ rather than the
`light-cone' one.

Finally, we consider an interplay between perturbative and
non-perturbative effects which becomes especially non-trivial in the
end-point domain. The Sudakov exponent introduces its own
smearing which is to be disentangled from that due to the primordial
motion. We have demonstrated how one can do this disentangling in
the radiative transitions and in the double spectral distributions in
the semi-leptonic decays.

\section{Limits on the Average Kinetic Energy of the Heavy Quark}

This section is related to the contents of the previous part in a
somewhat indirect manner. We will consider here general bounds on
the hadronic matrix elements of the operators built from heavy
quarks and some number of $\pi_\mu$'s. The first non-trivial
operator of this type is
\beq
{\cal O}_\pi =\bar Q {\vec\pi}^2 Q
\eeq
and we will mainly focus on discussion of ${\cal O}_\pi$. The matrix
element of ${\cal O}_\pi$ over $H_Q$ is parametrized as
\beq
\frac{1}{2M_{H_Q}}\langle H_Q|\bar Q {\vec\pi}^2 Q|H_Q\rangle
\equiv \mu_\pi^2 .
\eeq
Below we will need also the matrix element of the operator
${\cal O}_G$
\beq
{\cal O}_G =\bar Q \frac{i}{2}\sigma_{\mu\nu}G_{\mu\nu} Q,\,\,\,
\frac{1}{2M_{H_Q}}\langle H_Q|{\cal O}_G|H_Q\rangle
\equiv \mu_G^2 .
\eeq

It is clear that all operators of the type $\bar Q \pi_\mu\pi_\nu ... Q$
are related to the moments $a_n$ of the distribution function $F(x)$.
The positivity of the distribution function implies certain constraint
on the matrix elements of these operators, much in the same way as
the positivity of the light-quark distributions leads to constraints on
the matrix elements of the operators appearing in the OPE analysis of
deep inelastic scattering. These constraints are rather trivial.
Perhaps, the most well-known example in deep inelastic scattering is
the nucleon matrix element of the quark part of the
energy-momentum tensor. The positivity of the distribution
functions
implies that the quark's share of the nucleon momentum is always
less than one.  In the case of the heavy quarks the constraints are
weaker since the lower limit of variation of $x$ is unknown {\em
a priori}. Still, one can get an idea of the upper bound on
$\mu_\pi^2$
if one assumes  that $-x_0$ is close to $-1$.  There are absolutely no
good reasons to believe that $-x_0=-1$, still it is instructive to accept
this assumptions and see what the consequences are.

Were the function $F(x)$
symmetric in $x$ then we would automatically have $-x_0=-1$. We
could then easily obtain an upper bound on
$\mu_\pi^2$. Unfortunately, $F(x)$ is by no means symmetric. The
parameter $x_0$ introduced and discussed after eq. (\ref{31}) is not
equal to one. Moreover, perturbative effects enhance the tail at
negative $x$. At the same time inclusion of these effects (given all
our approximations and assumptions) does not change the fact that
$F(x)= 0$ at $x>1$. Thus, strictly speaking, we can {\em not} get a
rigorous
upper bound on $\mu_\pi^2$.
If for the purpose of orientation we
 assume for a moment that the distribution function is (nearly)
symmetric and $x_0 =1$ then $\int x^2dxF(x)< \int dx F(x) =1$
which implies, in turn, the bound
\beq
\mu_\pi^2\equiv\aver{\ve{\pi}^2} \le 3 \La^2 .
\label{S29}
\eeq
It is rather curious that the
 estimate of $\mu_\pi^2$  from the
QCD sum rules \cite{braun} nearly
saturates this limit.

One can obtain an upper limit on the second moment, however, with
no further assumptions on $x_0$, provided the third moment
is known from independent sources.
For instance, one can use the fact that for all $x\leq 1$
one has $x^2\leq 1+x-x^3$ that implies the inequality
\beq
a_2\le 1-a_3 .
\label{S30}
\eeq
Here we take into account  that $a_1=0$.

For an estimate of $a_3$ we will rewrite eq. (\ref{38a}) in the
following form
\beq
a_3\approx -\frac{1}{6\La^3}\, \frac{1}{2M_B}
\left( -\frac{2g^2}{9}\right)
\langle B|\bar b\gamma_5 q|0\rangle
\langle 0|\bar q \gamma_5 b|B\rangle =
\label{S31}
\eeq
$$
-\frac{g^2}{54}\frac{f_B^2M_B}{\La^3}
$$
where $D_\mu G_{\mu 0}$ is substituted by the light quark current
(the equation of motion) and then factorization is applied for the
matrix element over the $B$ meson state. Eqs. (\ref{S30}) and
(\ref{S31}) imply that
\beq
\langle {\vec\pi}^2\rangle \leq 3\La^2 +\frac{g^2}{18}\,
\frac{f_B^2M_B}{\La} .
\label{S32}
\eeq
The extra term in eq. (\ref{S32}) is positive which is welcome in
connection with the prediction \cite{braun} for
$ \langle {\vec\pi}^2\rangle $ mentioned above. The numerical value
of the extra term due to the third moment is rather uncertain. If
$g^2/4\pi$ is set equal to 1 then $(g^2/18)f_B^2M_B\sim 0.1$
GeV$^3$. The increase in  the naive bound (\ref{S29})  is quite
modest.

Let us proceed now to the discussion of the {\em lower} bound the
matrix element of the
operator
${\cal O}_\pi$. Unlike the previous case a strict lower bound
(strict in the quantum-mechanical sense) exists.

The physical meaning of this matrix element is the
average
kinetic energy (more exactly, the spatial momentum squared) of the
heavy
quark $Q$ inside $H_Q$. This operator is spin-independent, and we
are
aware of no method of  extracting  its matrix elements from
phenomenology,
without invoking additional theoretical arguments or models.

It is convenient to introduce a parameter similar to $\mu_G^2$
associated
with ${\cal O}_\pi$,
\begin{equation}
\frac{1}{2M_H} \langle H_Q|\bar Q {\vec\pi}^2 Q |H_Q \rangle
=\mu_\pi^2.
\label{62}
\end{equation}
It is plausible that $\mu_\pi^2$ for mesons and baryons is different
--
there is no reason why they should coincide.

For mesons a reliable lower
bound on $\mu_\pi^2$ can be given from essentially a quantum
mechanical
consideration. Indeed, let us take into account the fact that
\begin{equation}
[\pi_k\pi_l] =-i\epsilon_{kln}B_n
\label{63}
\end{equation}
where $B_n=-\epsilon_{npq}G_{pq}/2$ is a chromomagnetic field.
Then eq.(\ref{63}) immediately leads to
analog of the uncertainty principle for
$\pi_x$, $\pi_y$, namely
\beq
\langle \pi_x^2 \rangle \langle \pi_y^2 \rangle \geq
\frac {1}{4} \langle B_z \rangle^2
\label{64}
\eeq
where $\langle ... \rangle$ means average over some hadronic state.
It is convenient to use the polarized state of $B^*$  meson with
$S_z=1$
at this step. The average chromomagnetic field in this state can be
presented as a matrix element of the operator ${\cal O}_G$,
\beq
\langle B_z \rangle = - \frac {1}{2M_{B^*}}\langle
B^*(S_z=1)|{\cal O}_G |B^*(S_z=1) \rangle .
\label{65}
\eeq
Accounting for the fact that
the chromomagnetic field is proportional to
the spin of the light cloud, $\vec B = k {\vec S}_{lc}$,
one can relate the average (\ref{65}) to the average over the $B$
meson state ($S=0$)
and to the mass splitting between $B^*$ and $B$,
\beq
\langle B_z \rangle = \frac {1}{3} \, \frac {1}{2M_B}\langle
B|{\cal O}_G |B \rangle = \frac {\mu_G^2}{3}= \frac {1}{4}(M_{B^*}^2 -
M_B^2)
\label{66}
\eeq
It stems from three-dimensional rotational symmetry
in the rest frame  that
$$
\langle \pi_x^2 \rangle =\langle \pi_y^2 \rangle =\langle
\pi_z^2\rangle
$$
and, hence,
\begin{equation}
\langle {\vec \pi}^2 \rangle = \mu_{\pi}^2 (B)\geq
\frac{1}{2}\mu_G^2 .
(B)
\label{67}
\end{equation}

Numerically for B mesons $\mu_G^2\sim 0.35$ GeV$^2$ and then eq.
(\ref{67})
implies that $\mu_\pi^2\geq 0.18$ GeV$^2$.
Calculation \cite{braun} of this parameter from  the QCD sum rules
gives a value
which is only three times larger than our lower bound.  Namely,
according to \cite{braun}
$\mu_{\pi}^2 = 0.54\pm 0.12$ GeV$^2$.

Consideration of possible constraints  on $\mu_\pi^2$ from a
somewhat
different line of reasoning was  recently presented in \cite{neub}.

\section{Summary and Conclusions}

In this section we summarize our main results and findings. First and
foremost, it is demonstrated that  combining  QCD with the heavy
quark expansion provides us with a powerful tool for describing the
motion of the heavy quark inside the heavy hadrons. The formalism
emerging in this way  is model independent and similar to that used
in deep inelastic scattering. Technically our
consideration is based on the notion of the transition operator ${\hat
T}$ and  Wilsonian operator product expansion.  The corresponding
ideas can be traced to Refs. \cite{1,2,4,5}.

The inclusive spectra and the  decay rates in the semi-leptonic and
radiative transitions of the heavy flavor hadrons are determined by
universal distribution functions. The latter can be introduced through
matrix elements of the relevant local operators, much in the same
way it is routinely done in deep inelastic scattering.

The description of distributions in inclusive heavy flavor decays --
lepton or photon spectra in semileptonic or radiative beauty decays,
respectively --  obtained in QCD proper  would appear to be in
obvious
conflict with the data if the heavy quark $Q$ were at rest inside the
hadron $H_Q$. For in the simple case of radiative $B$ decays the
photon
spectrum would consist of a single line somewhat below the
kinematical
boundary. Intuitively it is obvious that $Q$ moves inside  $H_Q$.
The single photon line in radiative beauty decays then gets spread
out
by an amount of order $\La$. We have pointed out in the
past
\cite{5} and shown in explicit detail in the present paper that this
kind of Fermi motion -- rather than being forced upon us by
phenomenological constraints -- finds a natural home in a rigorous
QCD treatment. The salient features of the results obtained are:

$\bullet$ The motion of the heavy quark inside a given hadron is
described by a $universal$ function; i.e., it is the same function that
has to be folded with the primary lepton or photon spectra that arise
from the heavy quark decay to obtain the non-perturbative
corrections
to the observable spectra, in particular in the end-point region. This
universal function cannot be determined from perturbation theory --
like it is for the structure functions in DIS. The universality of the
heavy quark distributions take place only for a given value of the
final quark mass. Changing this mass we pass to a different
distrubition. In particular, two extreme cases have been considered
in some detail, the massless final quark and the non-relativistic
final quark. In the first case one has to deal with the `light-cone'
distribution function while in the second case it is the `temporal'
distribution function that is relevant.

$\bullet$ Although the motion of the heavy quark is non-relativistic
there exists $no$ non-relativistic hadronic wavefunction that can
reproduce the true effect of the Fermi motion beyond its second
moment
$\aver{(\vec p_F)^2}$.

$\bullet$ A model-independent lower bound has been derived for
$\aver{(\vec \pi)^2}$ that is not more than a factor of three below
what
has been inferred from QCD sum rules \cite{braun}.

$\bullet$ The impact of perturbative gluon emission on the spectra is
very considerable, in particular again in the end-point region.
Incorporating both perturbative as well as non-perturbative
corrections
in a self-consistent way poses a highly non-trivial problem. Our
treatment indicates that in beauty decays one can define and
combine
perturbative and non-perturbative corrections with sufficient
practical  precision.

\vspace{0.5cm}

{\bf ACKNOWLEDGMENTS:} \hspace{0.4cm} N.U. gratefully
acknowledges useful discussions with V. Braun and Yu. Dokshitzer.
We are greatful to L. Koyrakh for his generous assistance in
typesetting
the figures. This work
was supported in part by the National Science Foundation under the
grant number
PHY 92-13313 and by DOE under the grant number DOE-AC02-
83ER40105.

\vspace{0.5cm}
{\bf Note added.} \hspace{0.4cm} Two publications devoted to the
subject of the present work have just appeared \cite{falk,matthias}.
We plan to comment on the results of \cite{falk,matthias} in our
forthcoming paper \cite{BSUV2}.

\newpage

{\bf Figure captions}\\

\vspace{0.5cm}

Fig. 1. The spectrum in the transition $Q \ra \phi q $ for free
quarks.\\

Fig. 2. Smearing of the spectrum in $H_Q \ra \phi X_q $ due to the
heavy
quark primordial motion in $H_Q$.\\

Fig. 3. The graph determining the transition operator in the Born
approximation.\\

Fig. 4. The graph giving rise to the correction due to the operator
$\bar Q q \bar q Q$ in the transition operator.


\newpage

\begin{thebibliography}{99}

\bibitem{1}
N. Bili\'{c}, B. Guberina and J. Trampeti\'{c}, {\it Nucl. Phys.}
{\bf B248} (1984) 261;\\
M. Voloshin and M. Shifman, {it Yad. Fiz.} {\bf 41} (1985) 187
[{\it Sov. Journ.
Nucl. Phys.} {\bf 41} (1985) 120]; {\it ZhETF} {\bf 91} (1986) 1180
[{\it Sov. Phys. -- JETP} {\bf 64} (1986) 698].

\bibitem{2}
J. Chay, H. Georgi and B. Grinstein, {\it Phys. Lett.} {\bf B247} (1990)
399.

\bibitem{3}
E. Eichten and B. Hill, {\it Phys. Lett.} {\bf B234} (1990) 511;\\
H. Georgi, {\it Phys. Lett.} {\bf B240} (1990) 447;\\
for  recent reviews see e.g. \\
N. Isgur and  M.B. Wise, {\bf in:} {\em $B$
Decays},  S. Stone, Ed.  (World Scientific) 1992;\\
H. Georgi,  Preprint HUPT-91-A039 [Published in 1991 TASI
Proceedings]; \\
M. Neubert, Preprint SLAC-PUB-6263, 1993 [to be published in
{\it Phys. Reports}].

\bibitem{4}
I. Bigi, N. Uraltsev and A. Vainshtein, {\it Phys. Lett.} {\bf B293}
(1992)
430; (E) B297 (1993) 477;\\
B. Blok and M. Shifman, {\it Nucl. Phys.} {\bf B399} (1993) 441; 459;
\\
I. Bigi, B. Blok, M. Shifman, N. Uraltsev and A. Vainshtein, {\it Proc. of
the
1992 DPF meeting of APS}, Fermilab, November 1992  [Preprint
UND-HEP-92-BIG07].

\bibitem{5}
I. Bigi, M. Shifman, N. Uraltsev and A. Vainshtein, {\it Phys. Rev.
Lett.} {\bf 71} (1993) 496.

\bibitem{6}
B. Blok, L. Koyrakh, M. Shifman and A. Vainshtein, Preprint TPI-
MINN-93/33-T [Phys. Rev. D, to be published];\\
A. Manohar and M. Wise, Preprint UCSD/PTH 93-14;\\
T. Mannel, Preprint IKDA 93/26.

\bibitem{BU}
I.I. Bigi and  N.G. Uraltsev, {\it Phys. Lett.} {\bf 280B} (1992) 120.

\bibitem{bjorken}
J. D. Bjorken, {\it Phys. Rev.} {\bf D17} (1978) 171.

\bibitem{ACM}
G. Altarelli et al., {\it Nucl. Phys.} {\bf B208} (1982) 365.


\bibitem{randal2}
R. Jaffe and L. Randall, Preprint CTP-2189, 1993.

\bibitem{neubert}
M. Neubert, Preprint CERN-TH.7087/93

\bibitem{wilson}
K. Wilson, {\it Phys. Rev.} {\bf 179} (1969) 1499;\\
K. Wilson and J. Kogut, {\it Phys. Reports} {\bf 12} (1974) 75.

\bibitem{schwinger}
For comprehensive reviews see \\
V. Novikov, M. Shifman, A. Vainshtein and V. Zakharov,
{\it Fortschr. Phys.} {\bf 32} (1984) 585; \\
M. Shifman, Ed., {\em Vacuum Structure and QCD Sum Rules},
North-Holland, 1992, Chapter 3.

\bibitem{randal1}
G. Baillie, Preprint UCLA/93/TEP/47;\\
C. Csaki and L. Randall, Preprint CTP-2262, 1993.

\bibitem{BD}
G. Burdman and  J.F. Donoghue, {\it Phys. Lett.} {\bf 270B} (1991) 55.

\bibitem{chernyak}
V.A. Matveev, R.M. Muradyan and A.N. Tavkhelidze, {\it
Lett. Nuov. Cim.} {\bf 7} (1973) 719;\\
S. Brodsky and G. Farrar, {\it Phys. Rev. Lett.} {\bf 31} (1973)
1153;\\
V. Chernyak and A. Zhitnitsky, {\it Pis'ma ZhETF} {\bf 25} (1977) 544
[{\it JETP Lett.} {\bf 25} (1977) 510];\\
G. Farrar and D. Jackson, {\it Phys. Rev. Lett.} {\bf 43} (1979) 246;\\
G. Lepage and S. Brodsky, {\it Phys. Lett.} {\bf B87} (1979) 359;\\
A. Efremov and A. Radyushkin, {\it Phys. Lett.} {\bf B94} (1980)
245;\\
for an extensive review see\\
V. Chernyak and A. Zhitnitsky, {\it Phys. Reports} {\bf 112} (1984)
173.

\bibitem{shuryak}
E. Shuryak and A. Vainshtein, {\it Nucl. Phys.} {\bf B199} (1982) 451;
{\bf B201} (1982)  143;\\
R. Jaffe and M. Soldate, {\it Phys. Rev.} {\bf D26} (1982) 49.

\bibitem{SV}
M. Voloshin and M. Shifman, {\it Yad. Fiz.} {\bf 47} (1988) 801
[{\it Sov. J. Nucl. Phys.} {\bf 47} (1988) 511].

\bibitem{bjorken2}
J. Bjorken, Invited Talk at {\it Les Rencontres de la Valle d'Aosta,
La Thuille, 1990} Preprint SLAC-PUB-5278, 1990

\bibitem{J}
 Y. Azimov,  A. Vainshtein,  L. Lipatov and  V. Khoze,
{\it Pis'ma ZhETF} {\bf 21} (1975) 378 [{\it JETP Lett.}
{\bf 21} (1975) 172];\\
J. D. Jackson and D. L. Scharre,
{\it Nucl. Instrum. Meth.} {\bf 128} (1975) 13;\\
E. Kuraev and V. Fadin,
 {\it Yad. Fiz.} {\bf 41} (1985) 733
[{\it Sov. J. Nucl. Phys.} {\bf 41} (1985) 466].

\bibitem{DOK}
Yu.L. Dokshitzer, V.A. Khoze and S.I. Troian, Preprint LU TP 92-10,
1992.

\bibitem{novikov}
V. Novikov, M. Shifman, A. Vainshtein and V. Zakharov,
{\it Nucl. Phys.} {\bf B249} (1985) 445.

\bibitem{sh}
M. Shifman, Ed., {\em Vacuum Structure and QCD Sum Rules},
North-Holland, 1992.

\bibitem{BU4}
I.I. Bigi and N.G. Uraltsev, Preprint CERN-TH 7091/93 [to appear in
{\it Phys. Lett. B}].

\bibitem{braun}
P. Ball and V. Braun, Preprint MPI-Ph/93-51, 1993.

\bibitem{neub}
M. Neubert, Preprint CERN-TH.7070/93.

\bibitem{falk}
A. Falk, E. Jenkins, A. Manohar and M. Wise,
Preprint UCSD/PTH 93-38.

\bibitem{matthias}
M. Neubert,  Preprint CERN-TH.7113/93.

\bibitem{BSUV2}
I.I. Bigi, M. Shifman, N.G. Uraltsev and  A.I. Vainshtein, {\it in
preparation}.


\end{thebibliography}
\end{document}